\documentclass[twocolumn]{raa}
\usepackage{graphicx,times}
\usepackage{natbib}
\usepackage{amssymb,amsmath}
\usepackage{multicol}
\usepackage{multirow}
\usepackage[usenames,dvipsnames]{xcolor}
\bibpunct{(}{)}{;}{a}{}{,}

\newcommand{\aref}[1]{\hyperref[#1]{Appendix~\ref*{#1}}}

\usepackage{scalerel,tikz}
\usetikzlibrary{svg.path}
\definecolor{orcidlogocol}{HTML}{A6CE39}
\tikzset{orcidlogo/.pic={
 \fill[orcidlogocol] svg{M256,128c0,70.7-57.3,128-128,128C57.3,256,0,198.7,0,128C0,57.3,57.3,0,128,0C198.7,0,256,57.3,256,128z};
 \fill[white] svg{M86.3,186.2H70.9V79.1h15.4v48.4V186.2z}
 svg{M108.9,79.1h41.6c39.6,0,57,28.3,57,53.6c0,27.5-21.5,53.6-56.8,53.6h-41.8V79.1z M124.3,172.4h24.5c34.9,0,42.9-26.5,42.9-39.7c0-21.5-13.7-39.7-43.7-39.7h-23.7V172.4z}
 svg{M88.7,56.8c0,5.5-4.5,10.1-10.1,10.1c-5.6,0-10.1-4.6-10.1-10.1c0-5.6,4.5-10.1,10.1-10.1C84.2,46.7,88.7,51.3,88.7,56.8z};
}}

\newcommand\orcidicon[1]{\href{https://orcid.org/#1}{\mbox{\scalerel*{
\begin{tikzpicture}[yscale=-1,transform shape]
\pic{orcidlogo};
\end{tikzpicture}
}{|}}}}

\usepackage[pagebackref=true]{hyperref}

\hypersetup{colorlinks = true, linkcolor = blue, anchorcolor = red, citecolor = blue, filecolor = red, pagecolor = red, urlcolor = blue}

\begin{document}

   \title{A Study of Cataclysmic Variables from the eFEDS Survey}

 \volnopage{ {\bf 20XX} Vol.\ {\bf X} No. {\bf XX}, 000--000}
   \setcounter{page}{1}

\author{
Rui Wang\inst{1}
\and
Wei-Min Gu\inst{1}\thanks{Corresponding author: guwm@xmu.edu.cn}
\and 
Zhi-Xiang Zhang\inst{1}
\and 
Tuan Yi\inst{2}
\and
Senyu Qi\inst{1}
\and
Xiao-Jie Xu\inst{3}
}


   \institute{ Department of Astronomy, Xiamen University, Xiamen, Fujian 361005, China; {\it guwm@xmu.edu.cn}\\
        \and
             Department of Astronomy, School of Physics, Peking University, Beijing 100871, China\\
	\and
School of Astronomy and Space Science and Key Laboratory of Modern Astronomy and Astrophysics, Nanjing University, Nanjing, 210093, China\\
   {\small Received 20XX Month Day; accepted 20XX Month Day}
}

\abstract{We present 17 cataclysmic variables (CVs) obtained from the crossmatch between the Sloan Digital Sky Survey (SDSS) and eROSITA Final Equatorial Depth Survey (eFEDS), including 8 known CVs before eFEDS and 9 identified from eFEDS. The photometric periods of four CVs are derived from the Zwicky Transient Facility (ZTF) and Catalina Real-Time Transient Survey (CRTS). We focus on two CVs, SDSS J084309.3$-$014858 and SDSS J093555.0+042916, and confirm that their photometric periods correspond to the orbital periods by fitting the radial velocity curves. Furthermore, by the combination of the Gaia distance, the spectral energy distribution, and the variations of $\mathrm{H}\mathrm{\alpha}$ emission lines, the masses of the white dwarf and the visible star can be well constrained.
}

   \authorrunning{Wang et al. }            
   \titlerunning{A Study of Cataclysmic Variables from the eFEDS Survey}  
   \maketitle

%
\section{Introduction}

Cataclysmic Variables (CVs) are a class of interacting binary systems where a white dwarf (WD) primary accretes matter from a secondary star, typically a main-sequence star or a red dwarf \citep{2001cvs..book.....H}. CVs are the most abundant class of accreting compact object binaries within the Milky Way and provide valuable insights into the mechanisms of accretion physics \citep{2014apa..book.....G}. They also provide important information on the evolution of binary systems, with implications extending to various binaries.

CVs can be broadly classified into magnetic and non-magnetic types. In magnetic CVs, such as polar (with a magnetic field strength of roughly 10–250 MG) and intermediate polar (1–10 MG), the presence of strong magnetic fields significantly affects the accretion process \citep{2008ApJ...672..524N,2020MNRAS.494.3799P}. The accretion disc is either largely truncated or completely prevented from forming in magnetic CVs, resulting in matter only being able to collide with the surface of the white dwarf along the magnetic field lines \citep{2021NatAs...5..648S}. In contrast, non-magnetic CVs feature an accretion disc around the white dwarf, where material from the companion star is transferred via the Roche lobe. These systems exhibit significant photometric variability, including nova and dwarf nova outbursts, driven by instabilities within the accretion disc.

The orbital period is the most extensively studied parameter in CVs and serves as a key indicator of their evolutionary processes. For long orbital periods ($P_{\mathrm{orb}} > 3$ h), the evolution of CVs is primarily dominated by magnetic braking. The magnetic braking induced by the companion star's stellar wind \citep{1981A&A...100L...7V} results in angular momentum loss for the system, leading to a reduction in orbital separation and orbital period. However, when the orbital period decreases to around $P_{\mathrm{orb}}$ = 3 h, the companion star becomes fully convective, causing a reconfiguration of its magnetic field \citep{2009A&A...496..787R,2010A&A...513L...7S,2016MNRAS.457.3867Z}. This weakens magnetic braking significantly, leading to the companion star shrinking away from the Roche lobe, which halts mass transfer \citep{2011ApJS..194...28K}. Consequently, systems with orbital periods between 2 h and 3 h evolve as detached binaries, a phase known as the \text{\textquotedblleft}period gap\text{\textquotedblright}, during which they lose angular momentum solely through gravitational radiation. As the orbital period continues to decrease to $P_{\mathrm{orb}}$ = 2 h, the companion star again fills its Roche lobe, resuming mass transfer and continuing its evolution as a CV system. The number of CVs with well-determined orbital periods, particularly those within the period gap, remains limited. Estimating the periods of more CV systems would significantly contribute to refining models of CV evolution, thereby enhancing our understanding of binary star evolution.

In addition to orbital periods, measuring the masses of white dwarfs and companion stars in CVs is another key area of focus. Traditionally, white dwarf masses are determined by modelling the eclipse shape of the primary star. However, this approach relies on certain assumptions and requires a well-defined eclipse shape \citep{2005MNRAS.364.1158F,2019MNRAS.486.5535M}. Recently, \citet{2022ApJ...928...26G} proposed an alternative method using ultraviolet (UV) spectroscopy. The white dwarf mass can be directly estimated by simultaneously modelling the white dwarf and the iron curtain in the UV spectra. However, this method depends on the white dwarf being visible in the UV, and challenges may arise in distinguishing the white dwarf’s spectrum from that of the accretion disc.

In this paper, we investigate four poorly studied non-magnetic CVs: SDSS J084303.5$-$014858 (hereafter J0843), SDSS J084641.0+021823 (hereafter J0846), SDSS J091248.2$-$000721 (hereafter J0912) and SDSS J093555.1+042915 (hereafter J0935). We discuss the sample and the selection criteria for target sources in section~\ref{sec:target}. Section~\ref{sec:result} presents our results, and in Sections~\ref{sec:discuss} and~\ref{sec:summary}, we provide a discussion and a summary, respectively.


\section{Sample And Target Selection}
\label{sec:target}
\subsection{eFEDS Catalogue}
The extended ROentgen Survey with an Imaging Telescope Array \citep[eROSITA;][]{2021A&A...647A...1P}, aboard the SpektrumRÖntgen-Gamma (SRG) mission \citep{2021A&A...656A.132S} aims to deliver sensitive X-ray imaging and spectroscopy over a broad field of view. The SRG mission includes a four-year uninterrupted all-sky survey program \citep[the eROSITA All-Sky Survey: eRASS;][]{2021A&A...647A...1P}, which is expected to detect millions of X-ray sources for the first time. To demonstrate its groundbreaking survey capabilities, the eROSITA Final Equatorial-Depth Survey \citep[eFEDS;][]{2022A&A...661A...3S} observes a contiguous area of 140 square degrees during the SRG calibration and performance verification phase. Centred at RA 136 and Dec +2, it achieves a depth of about 2.2 ks ($\sim$ 1.2 ks after vignetting correction), with a limiting flux of $F_{\mathrm{0.5-2~keV}} \sim 6.5 \times 10^{-15} \text{erg s}^{-1} \text{cm}^{-2}$.

We use eFEDS point source counterpart catalogue, which includes both the \texttt{Main} and \texttt{Hard} catalogues. \citet{2022A&A...661A...3S} crossmatched eFEDS point-like sources with the DECam Legacy Survey (DECaLS) LS8 catalogue \citep{2019AJ....157..168D} to identify optical counterparts for the eFEDS X-ray sources. In this paper, we use the coordinates of the LS8 optical counterpart to crossmatch with the optical data, using a matching radius of 10 arcseconds. 

\begin{table}
\bc
\begin{minipage}[]{100mm}
\caption[]{Basic information on the 17 CVs.\label{table:2}}\end{minipage}
\setlength{\tabcolsep}{1pt}
\small
 \begin{tabular}{cccccccc}
  \hline\noalign{\smallskip}
SDSS ID & R.A. & Decl  & G mag & distance & $L_{\rm{X}}$ & $P_{\rm{orb}}$ & Ref \\
 & deg (J2000) & deg (J2000) &  mag  & pc & $\text{erg s}^{-1}$ & days \\
  \hline\noalign{\smallskip}
SDSS J084041.4+000520 & 130.17253 & 0.08894 &  20.96$\pm$ 0.03 &$-$ &$-$ &$-$ &(1)\\
SDSS J084303.5$-$014858 & 130.76455 & -1.81625  & 17.79$\pm$ 0.01 &$2045 ^{+555}_{-414}$& $6.0 \times 10^{31}$ & 0.3461&(9),(10)\\
SDSS J084400.1+023919 & 131.00042 & 2.65536  & 18.42$\pm$ 0.02 & $1467 ^{+674}_{-344}$ & $9.8 \times 10^{30}$ &0.2071&(2),(4)\\
SDSS J084641.0+021823 & 131.67094 & 2.30642 & 18.03$\pm$ 0.01 & $3577 ^{+799}_{-799}$ & $3.2 \times 10^{31}$&0.6566*&(9),(10)\\
SDSS J084735.4+014533 & 131.89749 & 1.75940 & $-$  &  $-$  & $-$ &$-$ &(9),(10)\\
SDSS J085107.4+030834 & 132.78081 & 3.14288 & 18.54$\pm$ 0.02 & $579 ^{+35}_{-35}$ & $4.0 \times 10^{30}$ &0.0665&(4)\\
SDSS J085300.6+020224 & 133.25236 & 2.07349 & $-$   & $-$  & $-$ &$-$&(9),(10)\\
SDSS J085550.9$-$015428 & 133.96189 & -1.90794  & 20.78$\pm$ 0.02 & $247 ^{+142}_{-82}$ & $2.5 \times 10^{29}$ &$-$&(9),(10)\\
SDSS J090246.5$-$014201 & 135.6939 & -1.70053  & 20.59$\pm$ 0.02 & $1735 ^{+1101}_{-667}$ & $1.8 \times 10^{31}$ &0.0850&(4)\\
SDSS J091248.2$-$000721 & 138.201 & -0.12259  & 20.61$\pm$ 0.01 & $995 ^{+419}_{-419}$ &  $1.9 \times 10^{30}$&0.0862*&(9),(10)\\
SDSS J091410.7+013733 & 138.54462 & 1.62582  & 17.69$\pm$ 0.01 & $1016 ^{+138}_{-103}$ & $7.1 \times 10^{30}$ &0.2518&(5),(6)\\
SDSS J091818.5+043610 & 139.57729 & 4.60277 &  $-$   & $-$ & $-$ & $-$&(9),(10)\\
SDSS J092614.3+010557 & 141.55955 & 1.09929  & 19.47$\pm$ 0.01 & $385 ^{+45}_{-41}$ & $4.9 \times 10^{30}$ &0.0613&(7)\\
SDSS J092902.9+005334 & 142.26193 & 0.89277 & $-$  &  $-$ & $-$ & $-$ &(9),(10)\\
SDSS J093205.2+034332 & 143.02179 & 3.72567  & 17.74$\pm$ 0.01 & $633 ^{+43}_{-46}$ & $1.9 \times 10^{30}$ &$-$&(6),(8)\\
SDSS J093238.2+010902 & 143.15922 & 1.15069  & 19.40$\pm$ 0.02 & $3145 ^{+1300}_{-1259}$ & $2.7 \times 10^{32}$ &$-$&(3),(4)\\
SDSS J093555.1+042915 & 143.97952 & 4.48787  & 18.65$\pm$ 0.01 & $364 ^{+29}_{-26}$ &   $1.06 \times 10^{30}$  & 0.1584 &(9),(10)\\
  \noalign{\smallskip}\hline
\end{tabular}
\ec
\tablecomments{\textwidth}{Column (1): SDSS ID of the source; column (2): R.A. (SDSS); column (3): Decl.(SDSS);  column (4): g-band magnitude from Gaia DR3; column (5): distance from \citet{2021yCat.1352....0B};  column (6): X-ray luminosity (0.2-2.3 keV); column (7): orbital periods, * represents a possible period; (8) References: (1) \citet{2014MNRAS.441.1186D}, (2) \citet{2003AJ....126.1499S}, (3) \citet{2012AJ....144...81T}, (4) \citet{2023MNRAS.524.4867I,2023MNRAS.525.3597I}, (5) \citet{2023AJ....165..148H}, (6) \citet{2023AJ....165..163C}, (7) \citet{2023ApJ...945..141R}, (8) \citet{2022MNRAS.516.2455N}, (9) \citet{2024A&A...686A.110S},(10) this work}
\end{table}

\begin{figure} 
   \centering
   \includegraphics[width=7.0cm, angle=0]{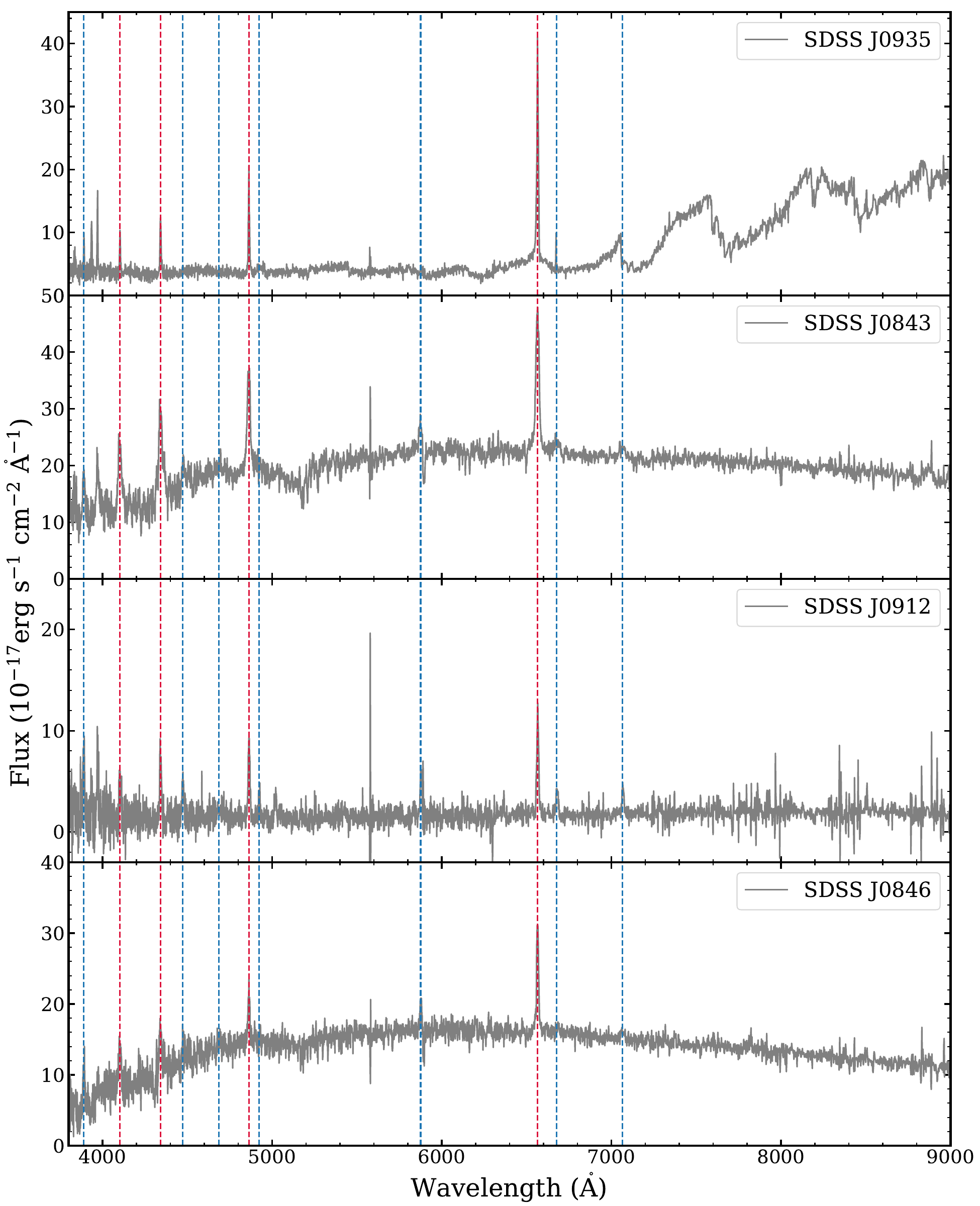}
   \includegraphics[width=7.0cm, angle=0]{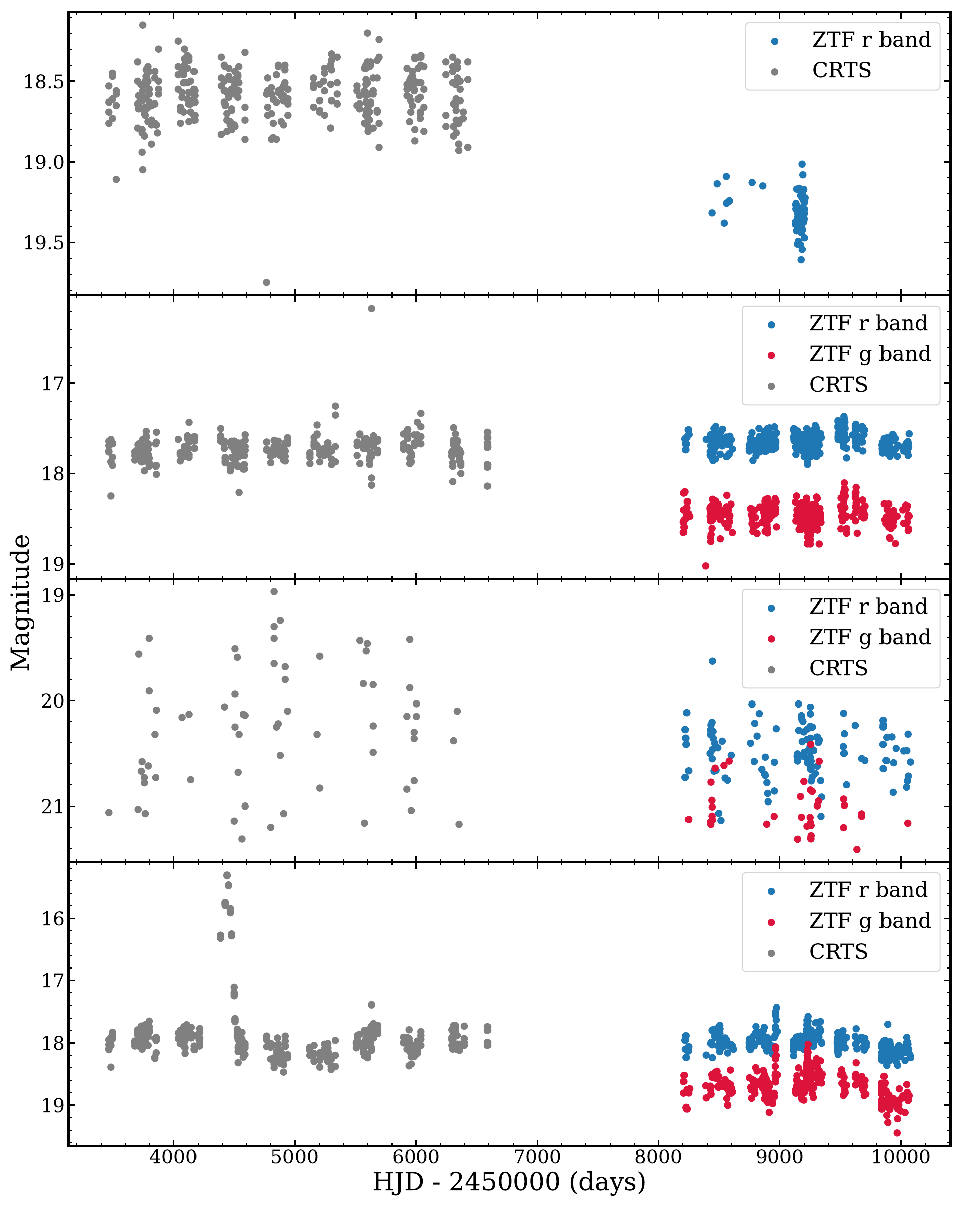}
\caption{Spectra and light curves of four CVs. The vertical lines in the left-hand panels indicate the rest wavelengths of the hydrogen Balmer lines (red dashed lines) and the helium \uppercase\expandafter{\romannumeral1} and \uppercase\expandafter{\romannumeral2} lines (blue dashed lines). Light curves with CRTS are shown in grey, ZTF in blue (g-band), and red (r-band) in the right-hand panels.}
\label{fig:fig1}
\end{figure}

\subsection{SDSS-V (SDSS DR18) Data}
The Sloan Digital Sky Survey (SDSS) is a major astronomical survey project that aims to provide comprehensive optical and infrared observations of the universe. Since its inception in 2000, SDSS has progressed through several phases  \citep{2000AJ....120.1579Y,2008AJ....135..338F,2009AJ....137.4377Y,2011AJ....142...72E,2017AJ....154...28B}, with each phase expanding its scope and incorporating new technological advancements. 

The current phase, SDSS-V \citep{2017arXiv171103234K}, continues this legacy by extending its coverage to the entire sky and offering multi-epoch optical and infrared spectroscopy. Specifically, SDSS-V focuses on eROSITA targets and has released spectra for 16,548 objects in the eFEDS field in its 18th data release \citep{2023ApJS..267...44A}. For our study, we crossmatch the SDSS-V DR18 data \footnote{\url{https://dr18.sdss.org/optical/spectrum/search}} with the sources from the aforementioned eFEDS catalogue, selecting only those sources classified as CLASS == STAR for our sample.

The SDSS BOSS spectrograph \citep{2013AJ....146...32S,2013AJ....145...10D} covers the range 3600 - 10400 $\rm\mathring{A}$. Each SDSS spectrum observation typically involves 2$-$4 consecutive exposures, spaced approximately 15 minutes apart, providing both individual exposures and combined spectra.  Figure~\ref{fig:fig1} shows the combined SDSS spectra for the four selected CVs. For each target, we have four or more individual exposure spectra, which are used to measure radial velocities (RVs). We limit our analysis to the 3600–9000 $\rm\mathring{A}$ range, where the Balmer line features are located and the absorption lines are of high quality.
\subsection{Targets Selection}
\label{subsec:target selection}
Due to the presence of accretion discs \citep{2008bhad.book.....K} or accretion columns \citep{2002apa..book.....F}, CVs exhibit distinct spectral features, including prominent Balmer and He emission lines. Accordingly, we plot the spectra of all target sources from the sample and visually inspect them, obtaining 17 CVs. These CVs are then crossmatched with the Simbad database to check for previously confirmed CVs and to obtain parameters such as  orbital periods. We find that 8 out of the 17 CVs have been already confirmed as CVs in Simbad \footnote{\url{http://simbad.u-strasbg.fr/simbad/}} before the release of the eFEDS catalogue, and the remaining 9 were identified from eFEDS and presented in a recent work \citep{2024A&A...686A.110S}. The basic information of these 17 CVs is listed in Table~\ref{table:2}. We subsequently examine the photometric data of the 9 eFEDS-identified CVs and find that four of them have usable photometric observations (with at least one filter containing more than 100 data points). The following sections will provide a detailed analysis of these four sources.

\begin{table}
\bc
\begin{minipage}[]{100mm}
\caption[]{Spectroscopic observational log.\label{table:1}}\end{minipage}
\setlength{\tabcolsep}{2.5pt}
\small
 \begin{tabular}{cccccccc}
  \hline\noalign{\smallskip}
ID & Telescope & HMJD & Exptime  & $\rm{RV}_{\rm{em}}$ & S/N (red)& $\rm{RV}_{\rm{abs}}$ & S/N (blue) \\
 &  &  & (s)  & ($\mathrm{km\,s^{-1}}$) &  & ($\mathrm{km\,s^{-1}}$)& \\
  \hline\noalign{\smallskip}

J0935 & SDSS 2.5-M & 59281.28339 & 900 & 5.02$\pm$ 3.65  &4.21& 19.54$\pm$ 2.5&11.67\\
J0935 & SDSS 2.5-M & 59281.29474 & 900 & 50.02$\pm$ 7.5 &4.31& 203.91$\pm$ 16&13.07\\
J0935 & SDSS 2.5-M & 59281.30608 & 900 & 36.06$\pm$ 9.2 &4.69& 240.39$\pm$ 20.4&11.87\\
J0935 & SDSS 2.5-M & 59281.31742 & 900 & 104.02$\pm$ 2.5 &3.84&257.37$\pm$ 42.5&9.47\\
J0935 & SDSS 2.5-M & 59284.26262 & 900 & -30.9$\pm$ 10.9 &4.31& -202.33$\pm$ 5&11.61\\
J0935 & SDSS 2.5-M & 59284.27398 & 900 & -11.94$\pm$ 1.0 &6.21&-180.85$\pm$ 1.2&12.53\\
J0935 & SDSS 2.5-M & 59284.28532 & 900 & -39.90$\pm$ 6.5 &4.95& -66.41$\pm$ 20.5&13.4\\
J0935 & SDSS 2.5-M & 59317.16325 & 900 & 105.00$\pm$ 10.5 &8.52& 195.91$\pm$ 3.0&18.62\\
J0935 & SDSS 2.5-M & 59317.17460 & 900 & 58.02$\pm$ 5.0 &8.65& 61.52$\pm$ 11.5&17.19\\
J0935 & SDSS 2.5-M & 59317.18594 & 900 & -11.94$\pm$ 2.0 &7.14& -74.91$\pm$ 10.0&16.65\\
J0843 & SDSS 2.5-M & 58931.13089 & 900 & -45.67$\pm$ 6.3 &11.20& 55.0$\pm$ 5.0&9.46\\
J0843 & SDSS 2.5-M & 58931.14227 & 900 & -40.53$\pm$ 10.5 &16.26& 12.13$\pm$ 5.0&10.00\\
J0843 & SDSS 2.5-M & 58931.15361 & 900 & -34.53$\pm$ 6.0 &14.00& -22.5$\pm$ 11.0&10.77\\
J0843 & SDSS 2.5-M & 58931.17632 & 900 & 26.33$\pm$ 6.5 &12.57& -124.1$\pm$ 11.0&7.83\\
  \noalign{\smallskip}\hline
\end{tabular}
\ec
\tablecomments{0.86\textwidth}{The HMJD refers to the heliocentric modified Julian date, and the RVs have also been corrected for the heliocentric frame. The $\rm{RV}_{\rm{em}}$ refers to RVs of the $\mathrm{H}\mathrm{\alpha}$ emission line. The $\rm{RV}_{\rm{abs}}$ refers to RVs of the absorption line.}
\end{table}

\section{Results}\label{sec:result}
The four targets selected in section~\ref{subsec:target selection} can be categorised based on their observations: two targets (J0843 and J0935) exhibit prominent stellar components in their spectra, while the other two (J0846 and J0912) do not. We will discuss these four targets below. Their basic information is presented in Table~\ref{table:2}. 

\subsection{Light Curves from ZTF and CRTS}
We obtain photometric data from the Catalina Real-Time Transient Survey \citep[CRTS;][]{2011arXiv1102.5004D} and Zwicky Transient Facility \citep[ZTF;][]{2019PASP..131a8002B}, which are shown on the right side of Figure~\ref{fig:fig1}. The light curves are used together with RVs to determine the orbital period.

\begin{figure}
\centering
\includegraphics[scale=0.27]{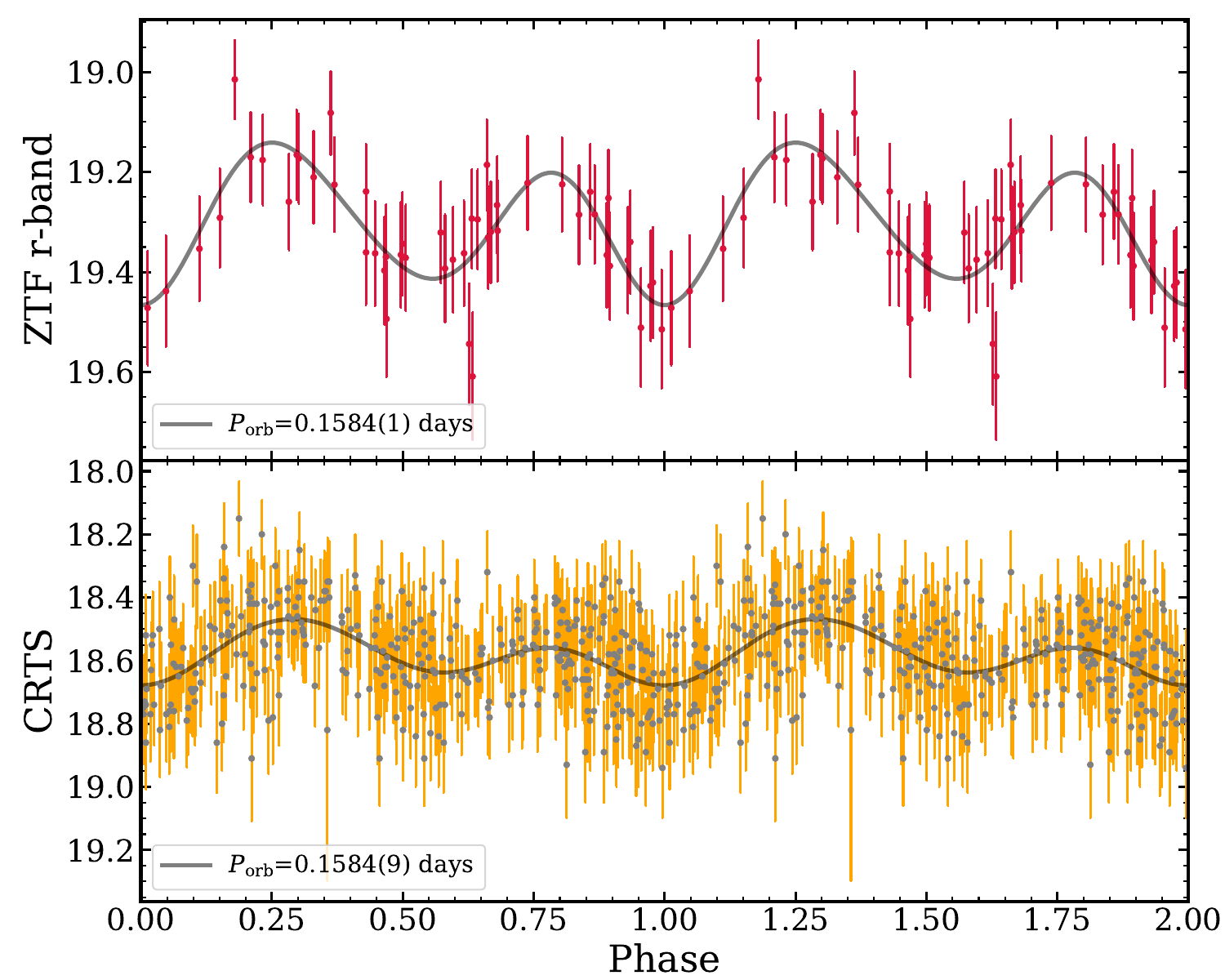}
\includegraphics[scale=0.27]{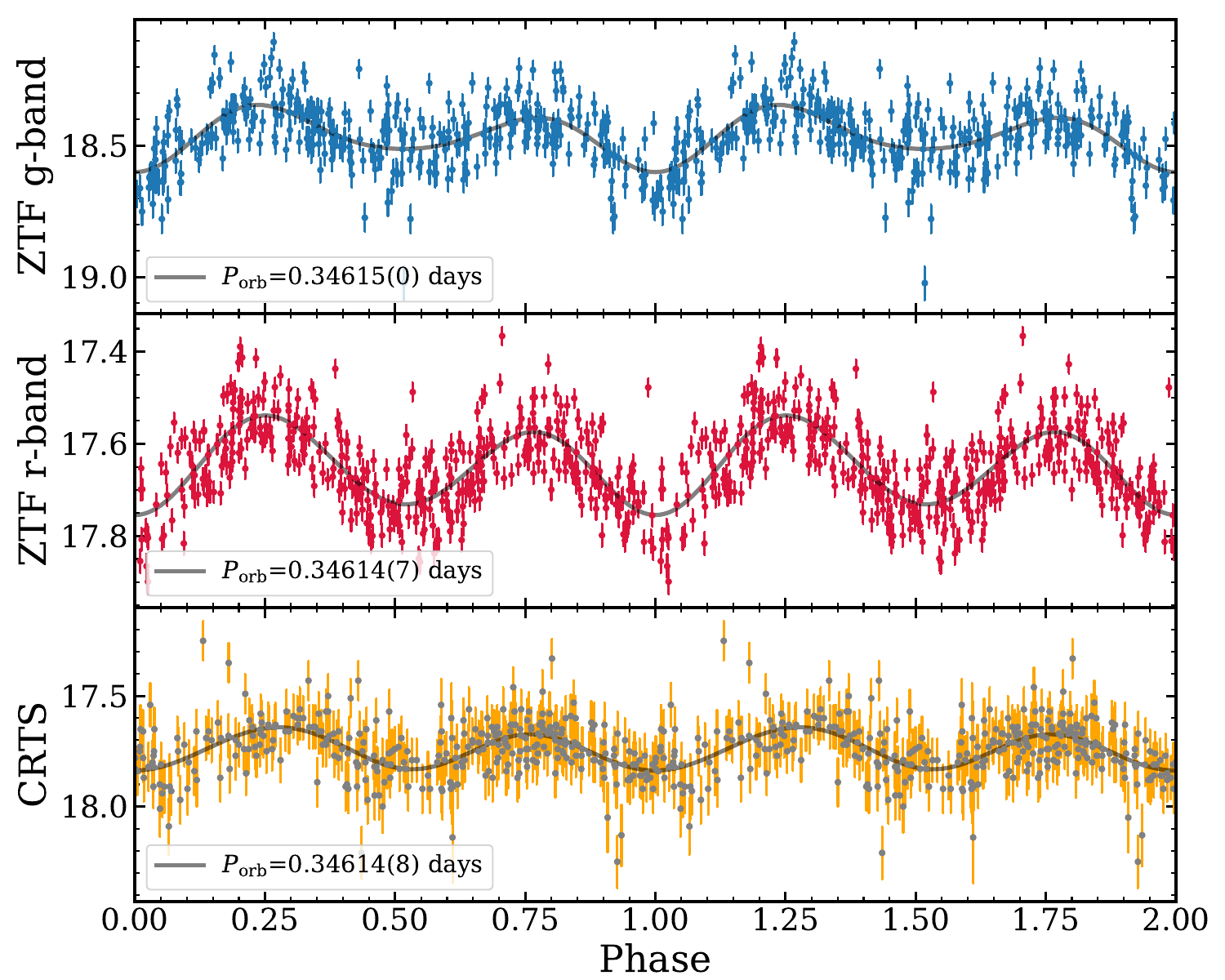}
\caption{The left panel shows the folded light curves of J0935 obtained from various surveys, and the right panel illustrates the folded light curves of J0843. The red and blue dots represent ZTF r-band and g-band data, respectively, while the yellow dots indicate CRTS data. The grey curves depict the three-harmonic model derived from Equation~(\ref{eq1}). The period of each light curve is denoted in the plot legend with uncertainty for the last digit shown inside the parentheses.}
\label{fig:fig2}
\end{figure}

We first use the light curves from ZTF/CRTS to search for photometric periods ($P_{\rm{ph}}$). We use the Lomb-Scargle periodogram \citep{1976Ap&SS..39..447L,1982ApJ...263..835S} to determine the photometric periods of the four targets. For J0843 and J0935, their photometric periods are 0.3461 days and 0.1584 days, respectively. Note that the periods are twice the peak periods provided by the Lomb-Scargle analysis due to their ellipsoidal modulation (a quasi-sinusoidal variation with double peaks and double valleys feature). We utilize a three-term Fourier model \citep{1993ApJ...419..344M}:
\begin{equation}\label{eq1}
\begin{split}
f(t) =& a_0\cos[\omega(t-T_0)]+a_1\cos[2\omega(t-T_0)]\\
&+a_2\cos[3\omega(t-T_0)]
\end{split}
\end{equation}
to fit the light curve and establish the zero point of the ephemeris $T_{0}$, where $\omega = 2\pi/P_{\mathrm{ph}}$, $a_{0}$, $a_{1}$, and $a_{2}$ are the fitting parameters. We minimize $\chi^{2}$ statistic to obtain the best-fitting parameters, which provides the zero point of ephemeris. Figure~\ref{fig:fig2} presents the folded light curves of J0843 and J0935 from various surveys or filters using $P_{\mathrm{ph}}$ and $T_{0}$. 

\begin{figure}
\centering
\includegraphics[scale=0.27]{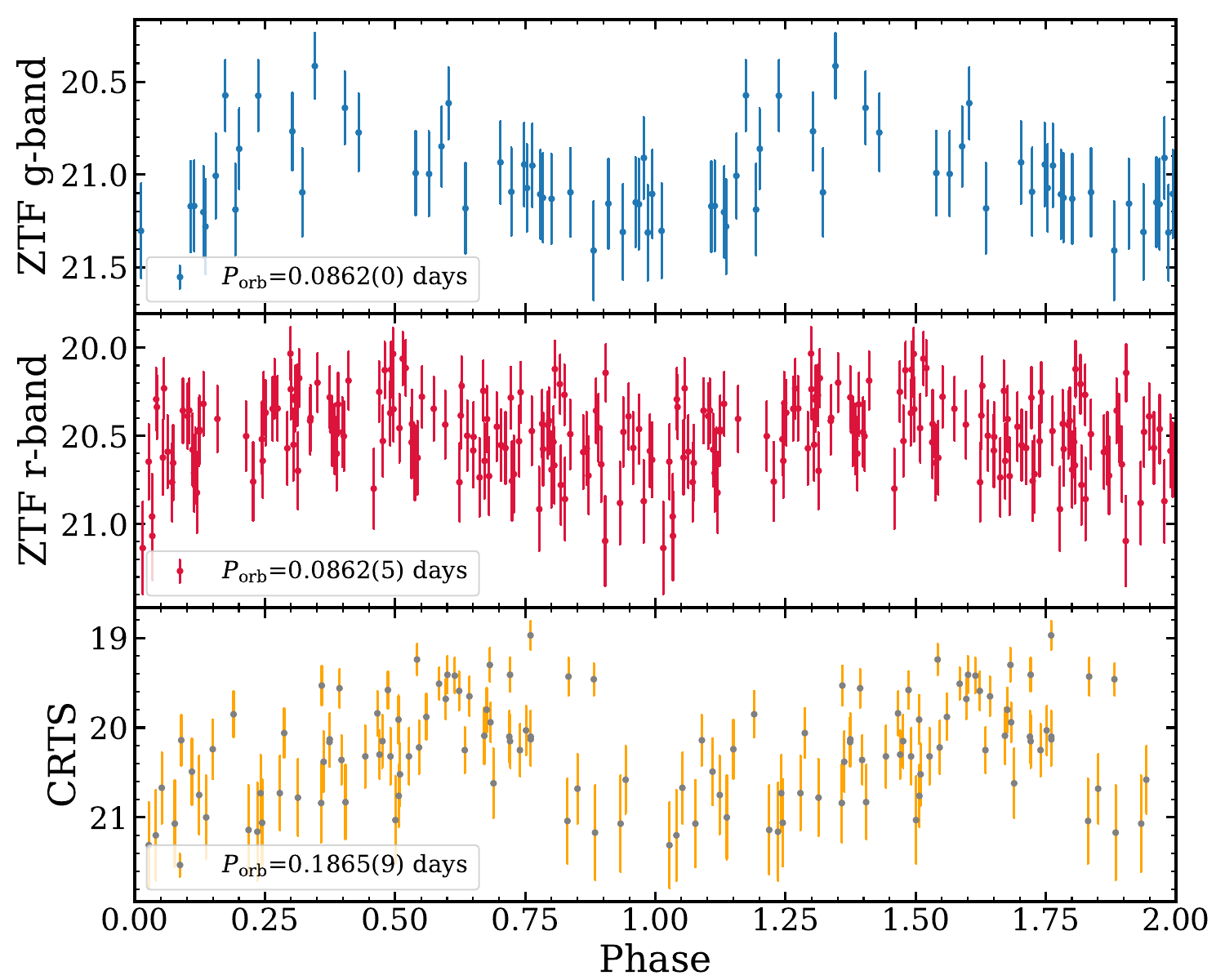}
\includegraphics[scale=0.27]{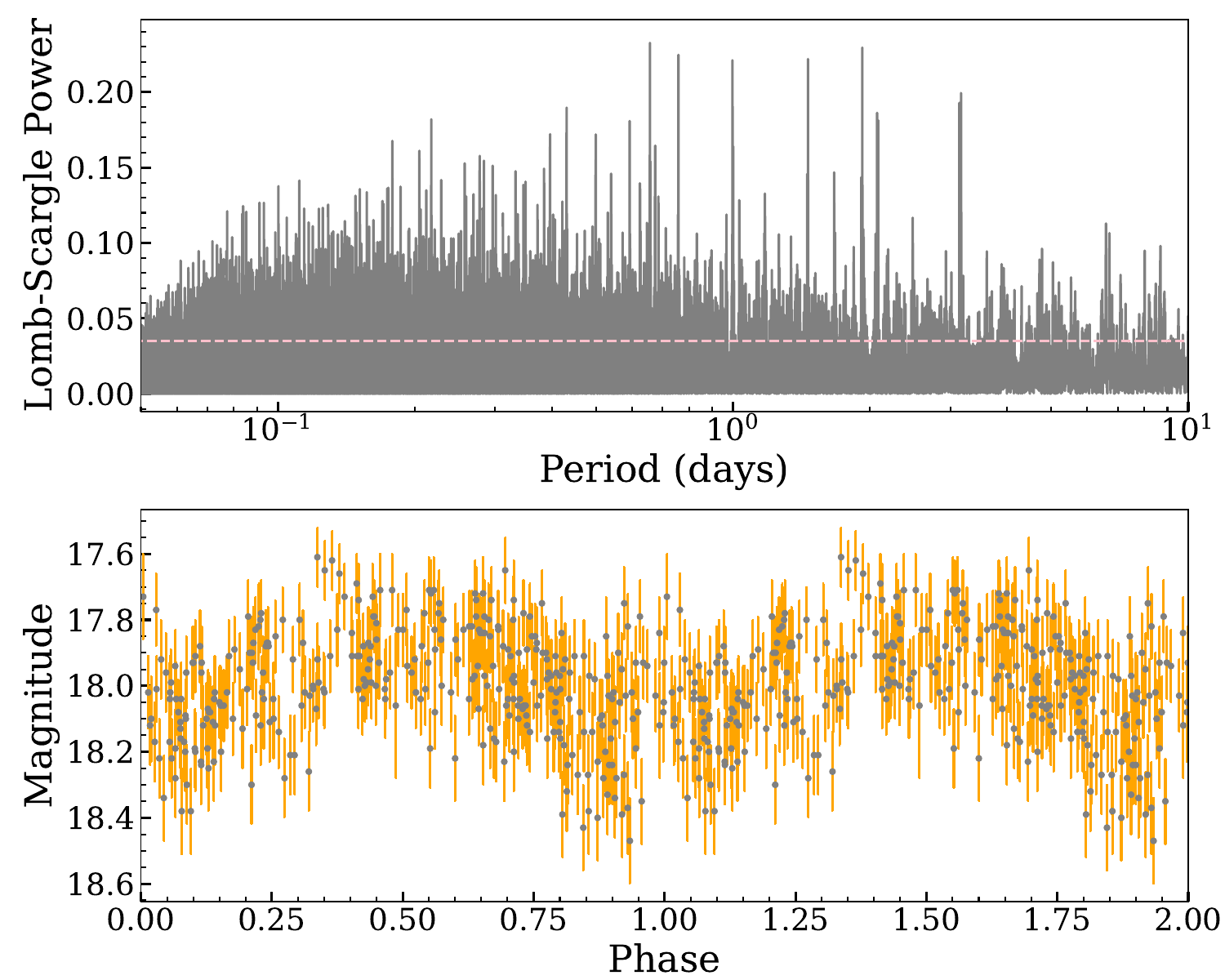}
\caption{The left panel: the folded light curves of J0912 obtained from various surveys. The right panel: the Lomb–Scargle periodogram and the phase-folded light curve of J0846 on the period of 0.6566 days. The red and blue dots represent ZTF r-band and g-band data, respectively, while the yellow dots indicate CRTS data.}
\label{fig:fig6}
\end{figure}

Figure~\ref{fig:fig6} illustrates the light curves of J0912 as obtained from various surveys. The periods identified from the ZTF g-band and r-band data are consistent but differ from those obtained from the CRTS data. This discrepancy could be due to J0912's relatively faint magnitude, which is close to the photometric limit of CRTS, leading to lower data quality. Therefore, the period of 0.0862 days provided by ZTF is considered the photometric period of J0912. J0846 exhibits a similar situation, where the period provided by ZTF is inconsistent with that from CRTS. However, the 0.4999 days given by ZTF are likely influenced by Earth's rotation. Figure~\ref{fig:fig6} shows the Lomb-Scargle power spectrum and the light curve of CRTS data for J0846, where the peak corresponding to 0.6566 days is not prominent. Therefore, additional photometric data may be necessary to confirm the photometric period of J0846.

We analyse the periods for the four CVs using light curves obtained from time-domain photometric surveys. The photometric periods of J0846 and J0912 exhibit some uncertainties due to the lack of prominent peaks in their Lomb-Scargle periodograms. In contrast, J0843 and J0935 display well-defined Lomb-Scargle peaks, allowing us to confidently identify their photometric periods as 0.3461 days and 0.1584 days, respectively. Next, we will fit the RVs of J0843 and J0935 to verify whether their photometric periods are associated with their orbital periods.

\subsection{Radial Velocity Curve of the Donor Star}
\label{subsec:donor rv}
\begin{figure}
\centering
\includegraphics[scale=0.30]{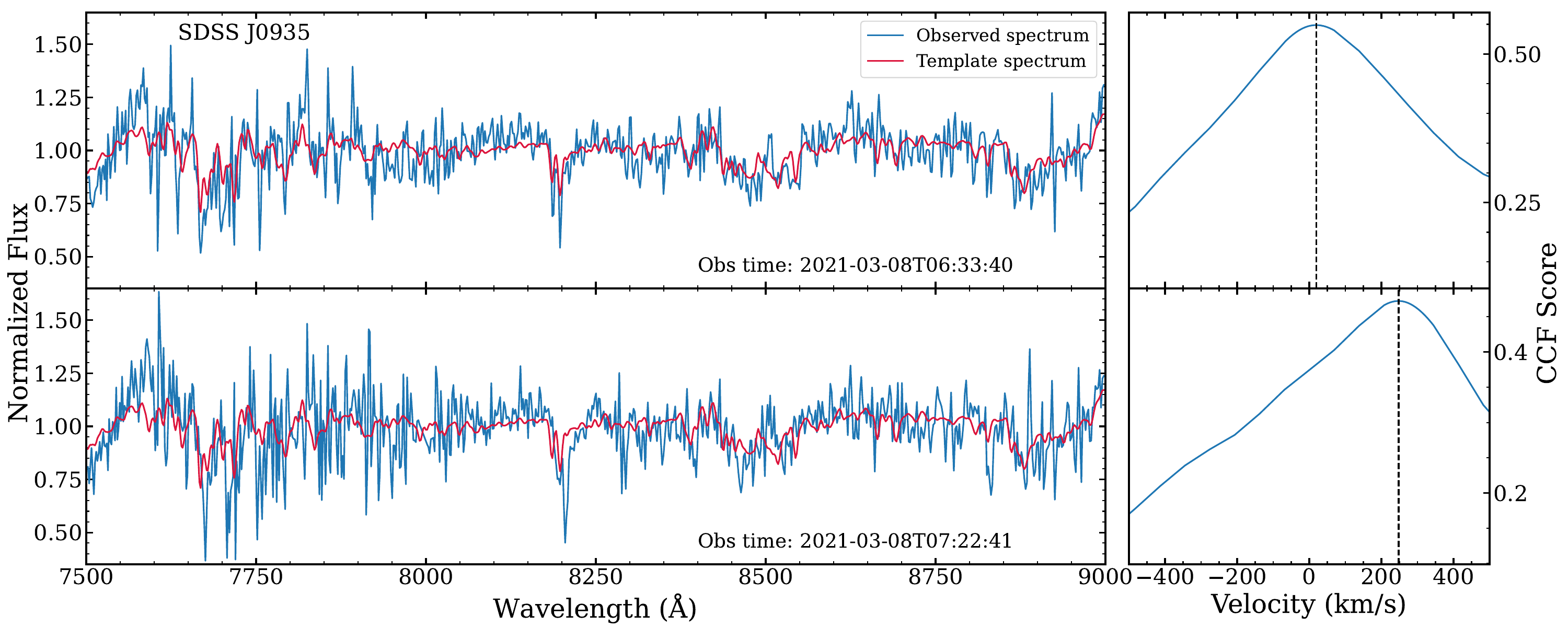}
\includegraphics[scale=0.30]{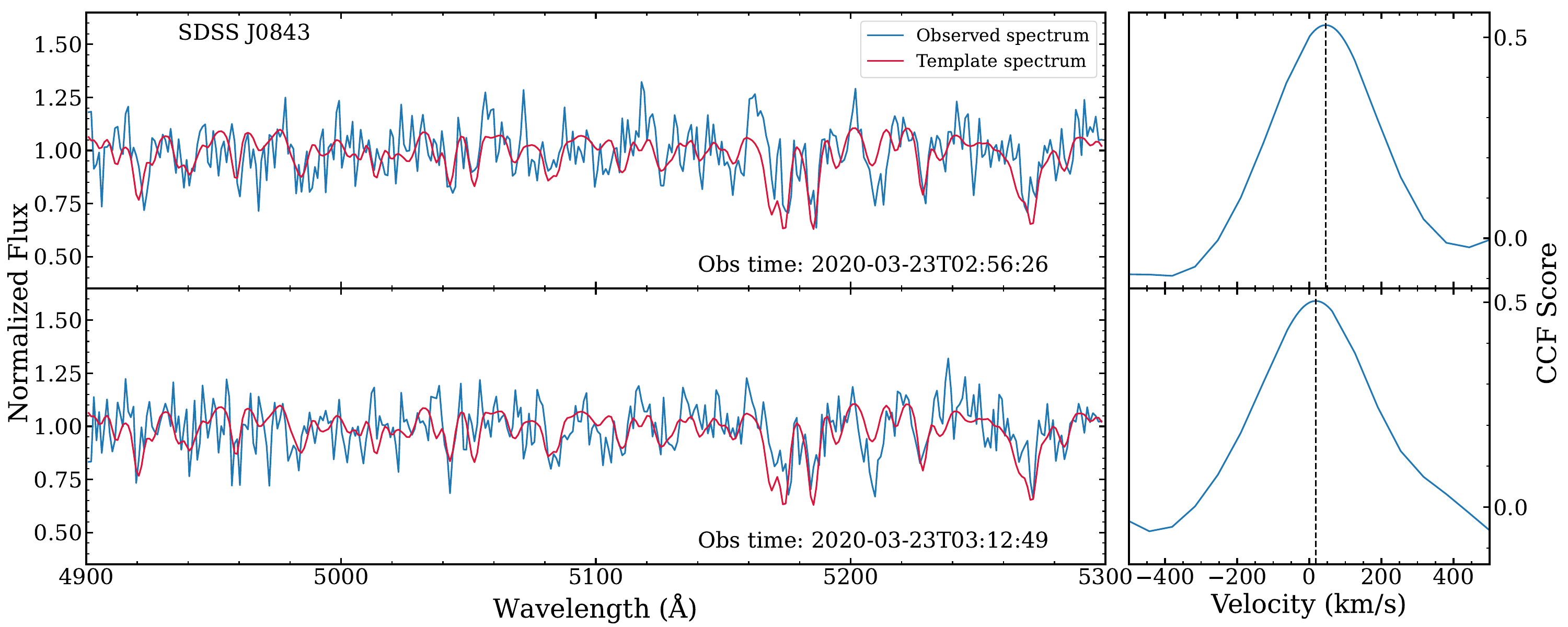}
\caption{SDSS spectra (left panels) and CCF (right panels) for J0935 (upper panels) and J0843 (lower panels). The left panels show the individual spectrum (blue) and a template spectrum (red) used to measure the CCF profile. The dashed lines mark the peaks in the CCF panels.}
\label{fig:fig3}
\end{figure}

In a binary system, when the visible star orbits an invisible companion, the spectral lines of the visible star exhibit periodic shifts due to the Doppler effect. These periodic shifts are reflected in the radial velocity of the visible companion \citep{2022NatAs...6.1203Y,2023AJ....165..187Q,2024ApJ...969..114L}. The spectra of J0843 and J0935 exhibit clear absorption features from their donor stars, allowing us to measure the donor star's RVs. We use the cross-correlation function (CCF) to measure the RVs from the SDSS spectra. The spectral wavelength region from 7000~$\rm\mathring{A}$ to 9000~$\rm\mathring{A}$ is used to measure RVs, excluding the disruption of Balmer and He \uppercase\expandafter{\romannumeral1}/\uppercase\expandafter{\romannumeral2} emission lines. Cosmic rays in the spectrum have been masked out. The optimal matching spectral template is selected by interpolating the Phoenix Models \citep{2013A&A...553A...6H}, a new library of high-resolution synthetic spectra based on the stellar atmosphere code PHOENIX \citep{1999JCoAM.109...41H}. We utilize the effective temperatures, log g and Fe/H values from Gaia DR3 for the donor star as prior information for template selection. Prior to matching, the model spectra are downgraded to roughly the same resolution as SDSS spectra by convolving with a Gaussian kernel. The best optimal stellar parameters are determined by minimizing the $\chi^{2}$ statistic between the observed spectra and the convolved templates. The left side of Figure~\ref{fig:fig3} displays the observed spectra of J0843 and J0935 alongside the best-matched template spectra. The profile of the CCF is shown on the right side of Figure~\ref{fig:fig3}. The observed and template spectra of J0843 and J0935 match well, and the single-peaked structure of their CCFs \citep{2017A&A...608A..95M} confirms that both are single-lined spectroscopic binaries.

To estimate the uncertainties in radial velocities, we apply the flux randomization/random subset sampling (FR/RSS) method \citep{1998ApJ...501...82P}. We generate 1000 mock spectra by perturbing flux values within their measurement uncertainties and randomly sampling subsets of the data. Each mock spectrum is then cross-correlated with the template spectrum to derive RVs, producing a distribution of mock RVs. The standard deviation of this distribution represents the uncertainty in RVs, accounting for both flux measurement errors and data point variability. The detailed RV values and errors are listed in Table~\ref{table:1}.

We assume that the visible star fills its Roche lobe, and the binary system follows a circular orbit. Therefore, we fit the RVs using a circular orbit model described by the equation:
\begin{equation}\label{eq2}
\begin{aligned}
V_{\mathrm{R}}=-K_{\mathrm{abs}} \sin[2\pi(t-T_{0})/P_{\mathrm{orb}}]+\gamma_{\mathrm{abs}}, 
\end{aligned}
\end{equation}
where $P_{\mathrm{orb}}$ is the orbital period estimated by the light curves, $K_{\mathrm{abs}}$ is the semi-amplitude of the visible star's RVs, and $\gamma_{\mathrm{abs}}$ is the systemic velocity. The photometric period corresponding to the peak in the Lomb-Scargle power spectrum does not accurately fit the RVs but fits well when doubled. We use 0.3461 days and 0.1584 days to fit the radial velocities of J0843 and J0935, respectively, and present their fitted RV curves in Figure~\ref{fig:fig5}. The best-fit results for the RVs show that the RV semi-amplitudes of the absorption lines for J0935 and J0843 are 263 $\pm$ 2 km s$^{-1}$ and 219 $\pm$ 7 km s$^{-1}$, respectively. The excellent agreement between the RV curves and the ellipsoidal light curve features confirms that the photometric periods of J0843 and J0935 are orbital periods.

For J0843, the orbital period matches the period inferred from the empirical relations for CVs described in \citet{2024A&A...686A.110S}. However, the orbital period of J0935 differs slightly from that reported by \citet{2024A&A...686A.110S}. They identify a period of 0.0733 days based on ZTF photometry, but our integrated analysis of photometric and spectroscopic RV fits suggests an orbital period of 0.1584 days for J0935. 

The spectra of J0846 and J0912, on the other hand, show no clear absorption features from their donor stars, containing only Balmer and He \uppercase\expandafter{\romannumeral1}/\uppercase\expandafter{\romannumeral2} emission lines, which prevents the measurement of the donor star's RV.

\subsection{Radial Velocity Curve of the Emission Line}
\label{RVEm}
Measuring the RV of the white dwarf directly is not feasible due to the absence of visible spectral features. In studies of CVs, previous methods for determining the white dwarf's mass have involved estimating its RV based on the observed motion of emission line profiles.

\begin{figure}
   \centering
   \includegraphics[scale=0.40]{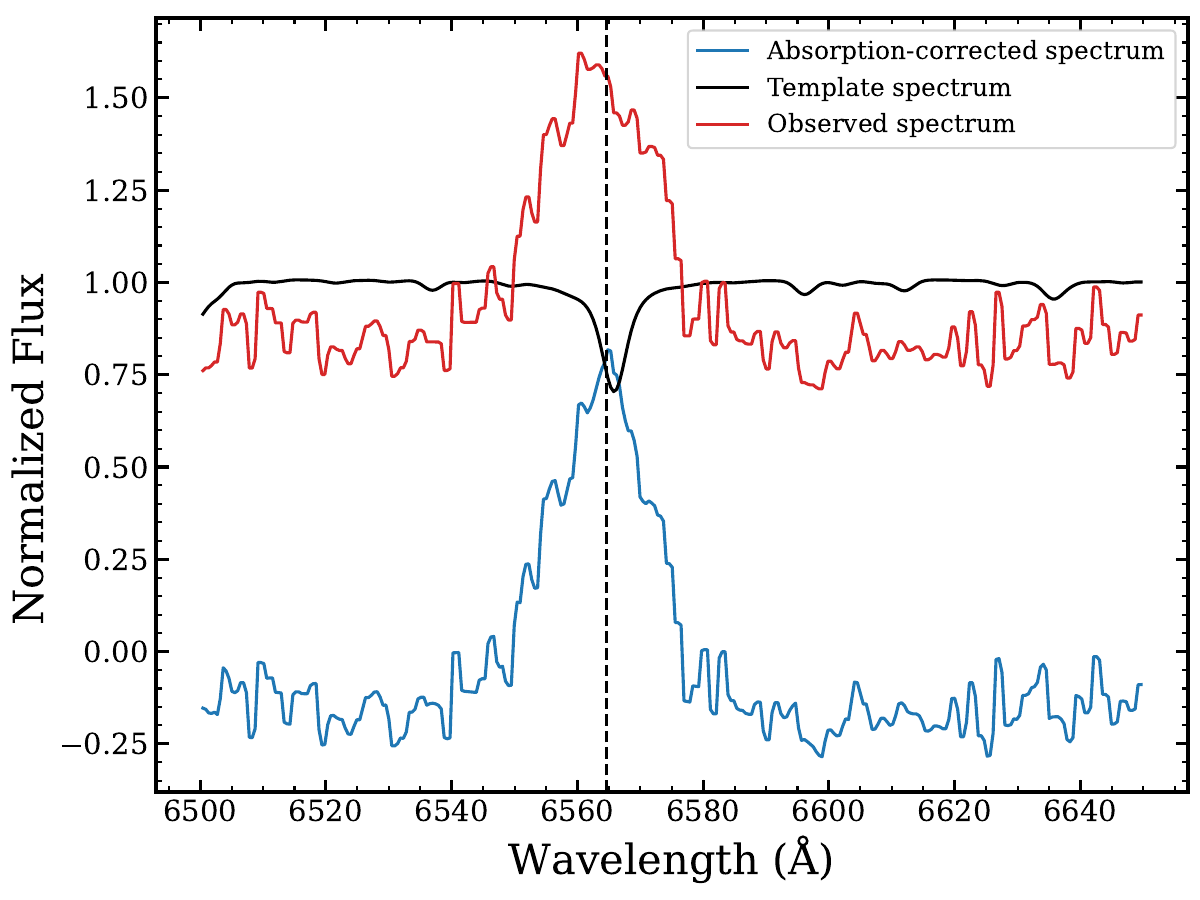}
   \includegraphics[scale=0.40]{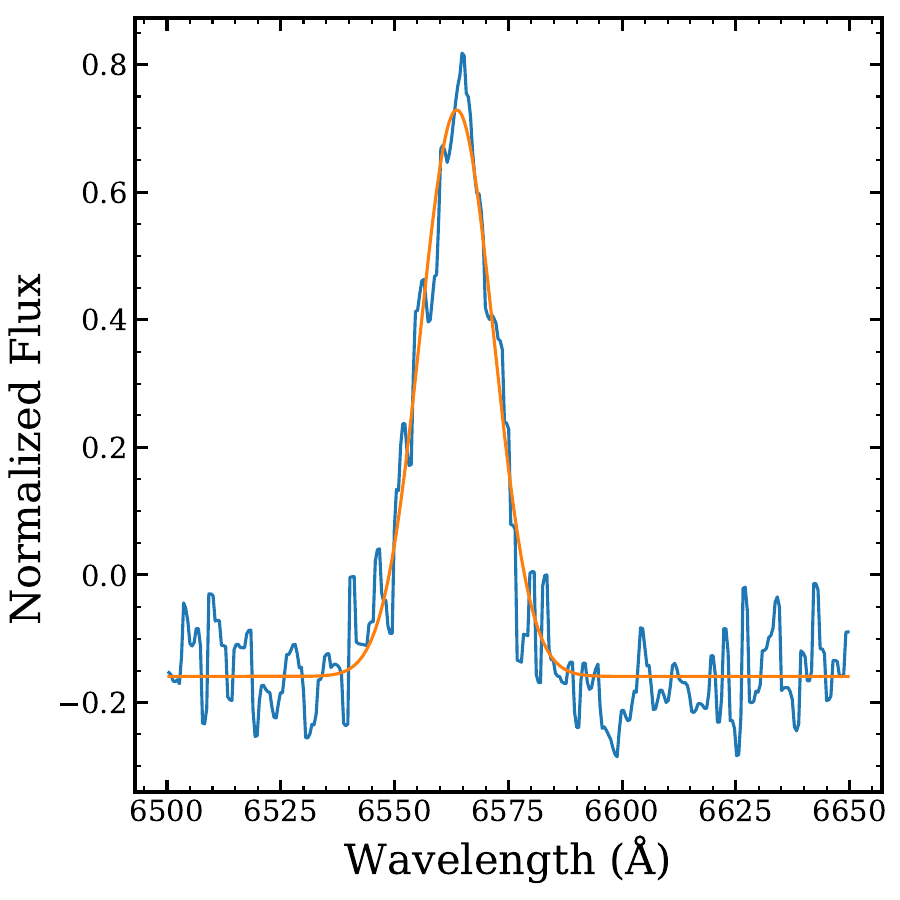}
      \caption{The measurement of the $\mathrm{H}\mathrm{\alpha}$ emission line center from one spectrum of J0935. In the left panel, the dashed line represents the reference rest wavelength of $\mathrm{H}\mathrm{\alpha}$ in a vacuum. In the right panel, the blue line shows the spectrum after subtracting the stellar component, while the yellow line represents the best Gaussian fit.}
         \label{fig:fig4}
   \end{figure}

To measure the radial velocities of the emission line, we follow a two-step process. In the first step, we select a single exposure spectrum with a high signal-to-noise ratio to measure the radial velocities of the emission lines using Gaussian fitting techniques. Based on the measured RVs of the absorption lines, the template spectrum selected in section~\ref{subsec:donor rv} is adjusted accordingly. This adjusted template spectrum is then subtracted from the single exposure spectrum, effectively removing the contributions of the absorption lines (as shown in Figure~\ref{fig:fig4}). We then apply Gaussian fitting to the absorption-corrected spectrum to determine the shift of the $\mathrm{H}\mathrm{\alpha}$ emission line centre, allowing us to derive the RV of the spectrum.

In the second step, the absorption-corrected spectrum from the first step is used as the template spectrum. CCF measurements are then performed between this template and the remaining single exposure spectra to determine their RVs. The method for estimating the RV errors is identical to that described in section~\ref{subsec:donor rv}. The RVs and their associated uncertainties are listed in Table~\ref{table:1}.

\begin{figure}
\centering
\includegraphics[scale=0.35]{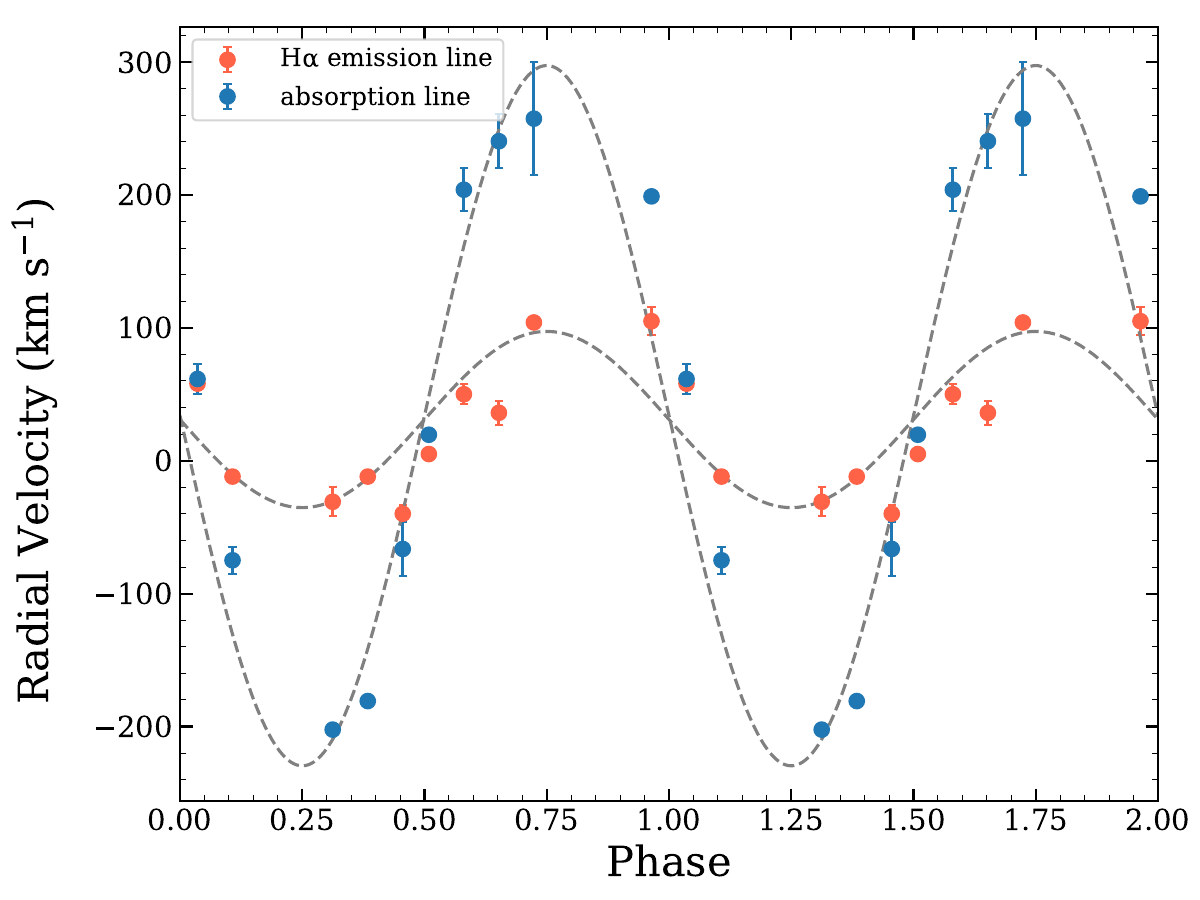}
\includegraphics[scale=0.35]{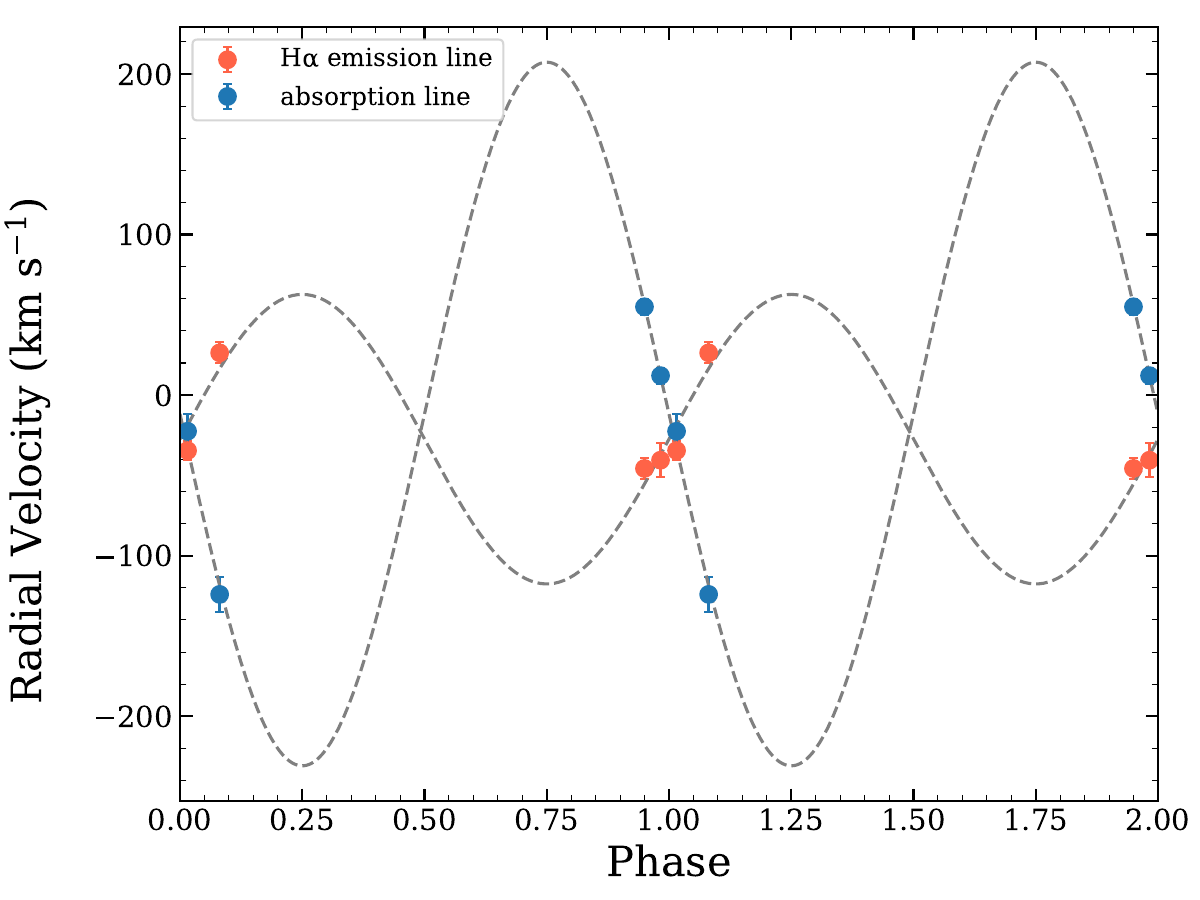}
\caption{Left: radial velocities of J0935 folded on the period of 0.15844 days. Right: radial velocities of J0843 folded on the period of 0.34615 days. The dashed lines depict the best-fitting RV curves. The blue dots represent RVs from absorption lines, while the RVs derived from $\mathrm{H}\mathrm{\alpha}$ emission lines are shown in orange.}
\label{fig:fig5}
\end{figure}

With the measured radial velocities in hand, we assume a circular orbit for the binary system and fit the data using the sinusoidal model $V_{\mathrm{R}}=-K_{\mathrm{em}} \sin[2\pi(t-T_{0})/P_{\mathrm{orb}}]+\gamma_{\mathrm{em}}$, where $K_{\mathrm{em}}$ is the semi-amplitude of RVs of the $\mathrm{H}\mathrm{\alpha}$ emission lines, $\gamma_{\mathrm{em}}$ is the systemic velocity. Fitting the RV data using twice the period corresponding to the peak also provides a good match. Figure~\ref{fig:fig5} displays the best-fit RV curves of J0843 and J0935. 

For J0843, the semi-amplitude of the RV curve of the emission lines is $90 \pm 6$ km s$^{-1}$. Given the opposite motion of the absorption and emission lines in J0843, we initially assume that the $\mathrm{H}\mathrm{\alpha}$ emission lines correspond to the motion of the white dwarf. Therefore, the mass ratio $q \equiv M_2/M_1$ is determined to be $0.41 \pm 0.03$ from the amplitude ratio of the emission and absorption lines.

For J0935, the RV curve of the emission lines follows the same direction as the absorption lines, with a semi-amplitudes of $K_{\mathrm{em}}$ = 66 $\pm$ 3 km s$^{-1}$. If we assume that the emission lines originate from the accretion disc and the absorption lines from the donor star, their RV curves should be opposite in direction, as observed in J0843. To explain this phenomenon, we analyse each exposure spectrum of J0935 and discover that the double-peaked structure of the $\mathrm{H}\mathrm{\alpha}$ emission line varies with the orbital phase. The most prominent peaks occur at phases approximately 0.25 and 0.75. These features most likely reflect the $\mathrm{H}\mathrm{\alpha}$ emission arising from irradiated regions on the M-dwarf companion, whose projected velocity varies with orbital phase \citep{2001ApJ...562L.145S,2004ApJ...614..338B,2019AJ....158..156G,2025A&A...698A..81L}. While a binary M-dwarf system may produce a similar pattern, we rule out this possibility in section~\ref{parameter}.

\subsection{Spectral Energy Distribution of the Donor Star}
\label{SED}
Due to the compact size of the white dwarf and the relatively low temperature of the secondary star, their contributions to the optical spectrum are minor compared to the dominant influence of the luminous accretion disc. However, as one moves from the optical to the optical-infrared range (0.7 to 1.0 \text{$\mu$m}), the relative contribution of the accretion disc diminishes, allowing spectral features from the red companion star to become visible \citep{1988MNRAS.233..451F}. Spectral energy distributions (SEDs) analysed by \citet{2002ApJ...569..395D} and \citet{2003ApJ...592.1124T} indicate that substantial circumbinary discs (CB) begin to surpass the emission from the binary system at approximately 5 \text{$\mu$m} in the mid-infrared range.
Considering the uncertainties in the distance and the possibility that secondary stars in CVs may be brighter than isolated main-sequence stars of the same type, \citet{2004MNRAS.349..869D} propose that no additional components beyond the secondary star are necessary for the near-infrared SED of CVs. The companion star components of J0843 and J0935 are visible in their spectra, indicating that the contribution from their companions is relatively significant. Therefore, we choose the wavelength range from 0.7 to 5 \text{$\mu$m} for SED fitting to derive the stellar properties of the secondary star. For simplicity, we assume that all fluxes within this wavelength range are solely contributed by the secondary star of the system, neglecting contributions from the accretion flow, hotspot, white dwarf, and circumbinary disc.

   \begin{figure}
   \centering
   \includegraphics[scale=0.25]{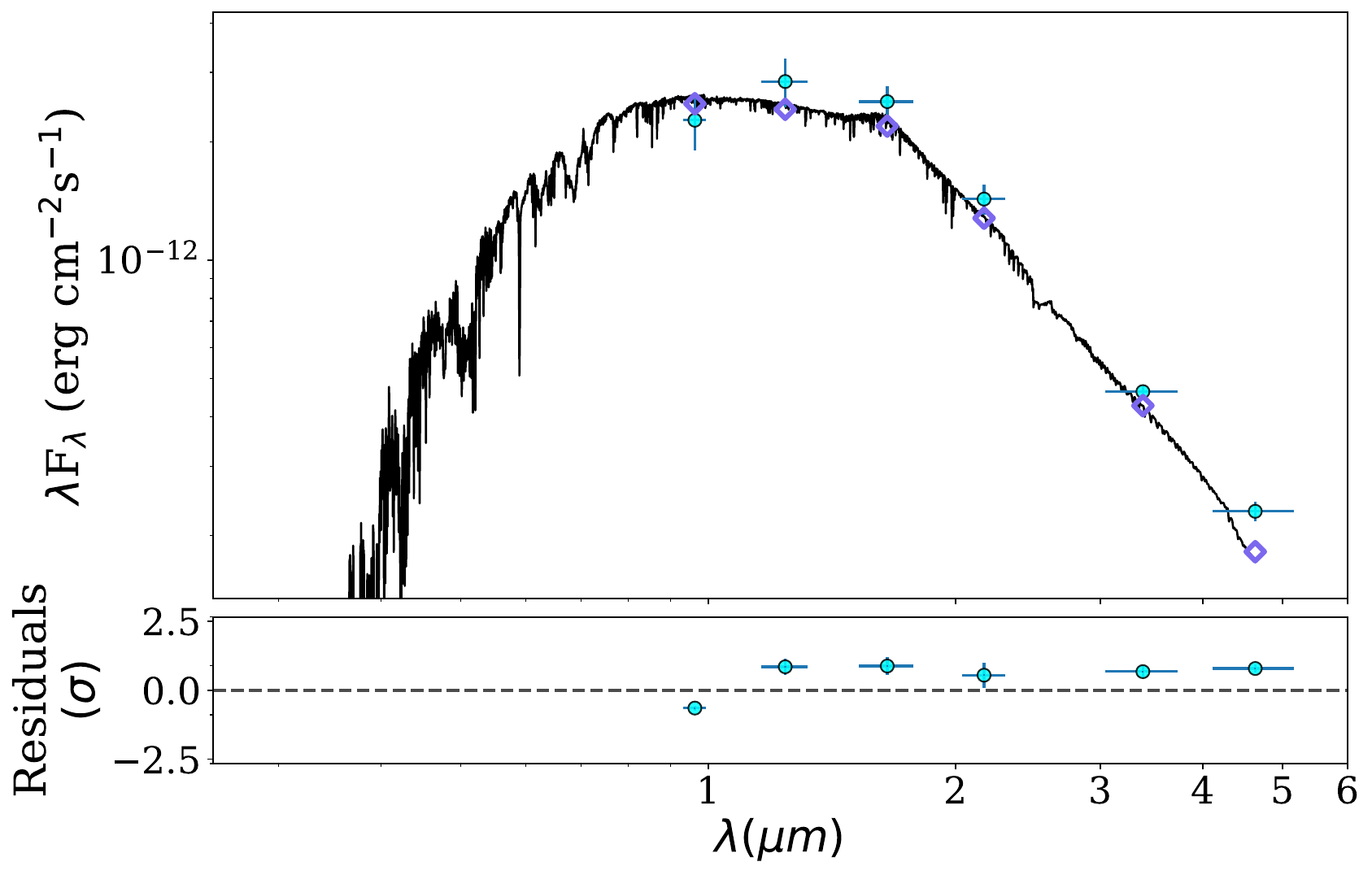}
   \includegraphics[scale=0.25]{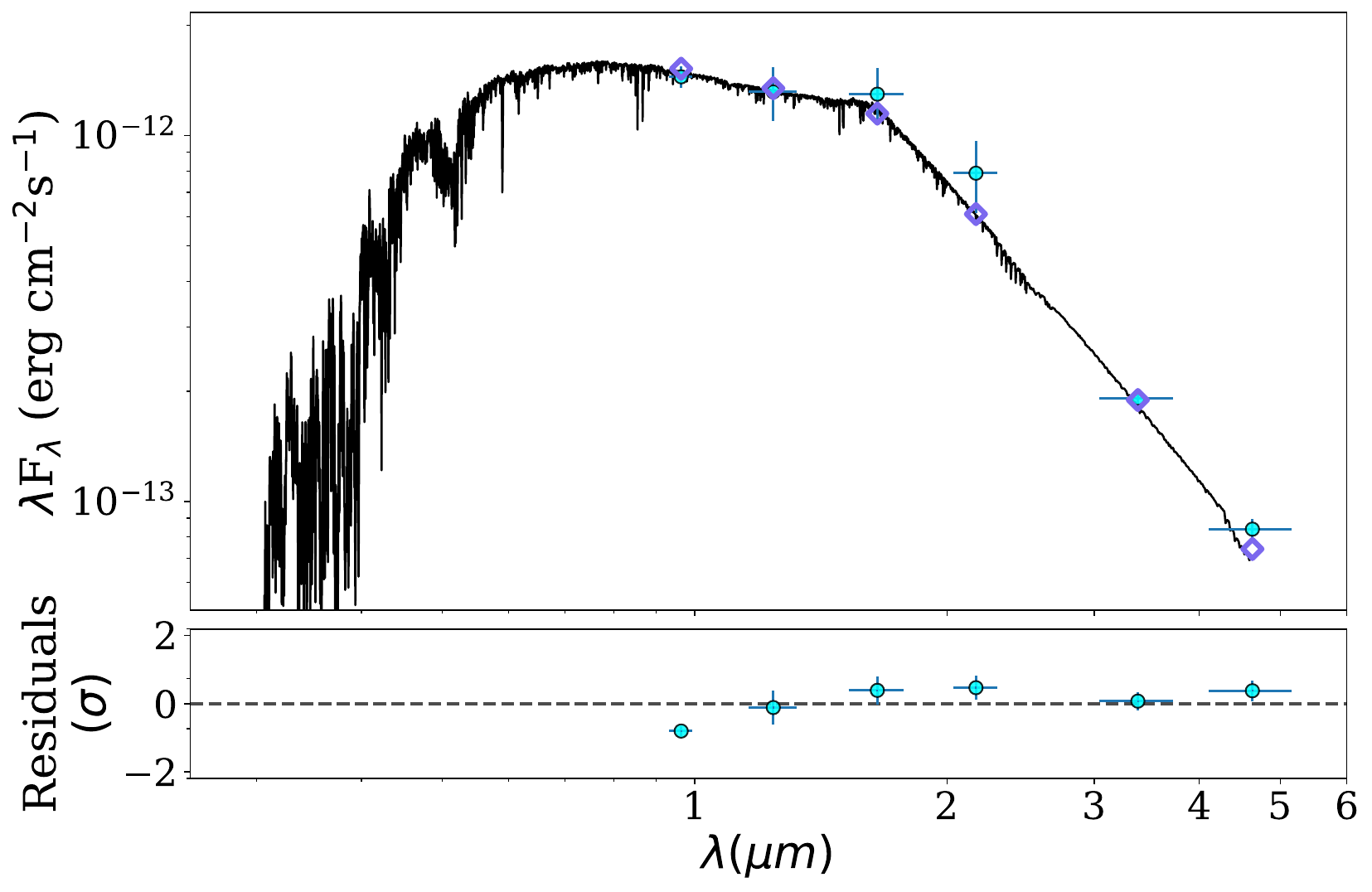}
      \caption{SED fitting results for J0935 (left panel) and J0843 (right panel). Blue points denote observed fluxes, purple diamonds indicate synthetic fluxes, and the black curve represents the best-fitting SED model.}
         \label{fig:fig7}
   \end{figure}

In order to constrain the stellar parameters, we fit SEDs of J0843 and J0935, both of which show a significant secondary star component in their spectra. For the SED fitting, we utilize the \texttt{astroARIADNE} \footnote{\url{https://github.com/jvines/astroARIADNE}} software package, which is designed to automatically fit stellar SEDs using up to six distinct atmospheric model grids, allowing us to determine the effective temperature, surface gravity, metallicity, distance, radius, and V-band extinction \citep{2022MNRAS.513.2719V}.

\texttt{AstroARIADNE} automatically collects multi-band photometric data from sources including GALEX \citep{2005ApJ...619L...1M,2011Ap&SS.335..161B}, SDSS \citep{2009ApJS..182..543A}, APASS \citep{2014CoSka..43..518H}, Pan-STARRS \citep{2016arXiv161205560C}, TESS \citep{2015JATIS...1a4003R}, 2MASS \citep{2006AJ....131.1163S} and ALLWISE \citep{2010AJ....140.1868W} for SED fitting. Given the influence of the white dwarf and the accretion disc in the ultraviolet and optical wavelength ranges, we limit the photometric data used for the SED fitting to Pan-STARRS (PS1), 2MASS, and WISE ($W_{1}$, $W_{2}$) only, where the contribution from the secondary star is more dominant. All the photometric data are summarized in Table~\ref{table:11}. We employ the distance measured by inverting the trigonometric parallax from Gaia DR3 \citep{2023A&A...674A...1G} as a prior for the distance parameter, and the $A(V)$ value from the 3D Dust Mapping \citep{2018JOSS....3..695M} as a prior for V-band extinction. Furthermore, according to the Chandrasekhar limit \citep{1931ApJ....74...81C}, the upper mass limit for white dwarfs is $1.4~M_{\odot}$. Based on the mass ratio, the visible star mass should also have an upper limit of $0.574~M_{\odot}$. Correspondingly, using the period–density relationship (Equation~(\ref{eq14})), the upper radius limit for the visible star is $0.797~R_{\odot}$, which is used as the prior for the radius parameter. The other parameters are treated as free parameters. The stellar parameters, including radius($R_{2}$), effective temperature ($T_{\rm{eff}}$), metallicity ([Fe/H]), and surface gravity (log~$g$) are derived by fitting the SEDs with atmospheric model grids. However, considering that CVs are accretion systems, we did not adopt the isochrone masses provided by stellar evolution models. All the best-fitting results are summarized in Table~\ref{table:parameters}, with the best-fit models shown in Figure~\ref{fig:fig7}.

\begin{table}
\bc
\begin{minipage}[]{100mm}
\caption[]{The SED of J0935 and J0843.\label{table:11}}\end{minipage}
\setlength{\tabcolsep}{1pt}
\small
 \begin{tabular}{cccccccc}
  \hline
\hline
Survey & Filter & $\lambda_\text{effective}$ & AB mag(J0843) & Vega mag(J0843) &log $\lambda f_{\lambda}$(J0843) \\
 &  & (\text{$\mu$m}) &(mag) & (mag) &log ($\text{erg s}^{-1} \text{cm}^{-2}$)&  \\
  \hline\noalign{\smallskip}
Pan-STARRS & PS1 y & 0.963 & 17.23$\pm$ 0.01 &  & -11.823$\pm$ 0.004 \\ \hline
      & 2MASS J & 1.241 &  & 16.17$\pm$ 0.11 & -11.974$\pm$ 0.045 \\
2MASS & 2MASS H & 1.651 &  & 15.40$\pm$ 0.13 & -12.104$\pm$ 0.054 \\
      & 2MASS $K_{\text{s}}$ & 2.166 &  & 15.17$\pm$ 0.19 & -12.438$\pm$ 0.076 \\ \hline
WISE & $W_{1}$ & 3.379 &  &15.41$\pm$ 0.04 & -13.247$\pm$ 0.016 \\
     & $W_{2}$ & 4.629 &  & 15.32$\pm$ 0.10& -13.741$\pm$ 0.041 \\ 
\hline
\hline
Survey & Filter & $\lambda_\text{effective}$ & AB mag(J0935) & Vega mag(J0935) &log $\lambda f_{\lambda}$(J0935) \\
 &  & (\text{$\mu$m}) &(mag) & (mag) &log ($\text{erg s}^{-1} \text{cm}^{-2}$)&  \\
\hline
Pan-STARRS & PS1 y & 0.963 & 16.74$\pm$ 0.01&  & -11.628$\pm$ 0.006 \\ \hline
      & 2MASS J & 1.241 &  & 15.33$\pm$ 0.06 & -11.640$\pm$ 0.023 \\
2MASS & 2MASS H & 1.651 &  & 14.68$\pm$ 0.05 & -11.815$\pm$ 0.021 \\
      & 2MASS $K_{\text{s}}$ & 2.166 &  & 14.53$\pm$ 0.10 & -12.180$\pm$ 0.039 \\ \hline
WISE & $W_{1}$ & 3.379 &  & 14.45$\pm$ 0.03 & -12.863$\pm$ 0.012\\
     & $W_{2}$ & 4.629 &  & 14.22$\pm$ 0.05 & -13.303$\pm$ 0.018 \\
  \noalign{\smallskip}\hline
\end{tabular}
\ec
\tablecomments{1.0\textwidth}{The AB and Vega magnitude systems are listed in two columns.}
\end{table}

\subsection{Binary Parameters}
\label{parameter}
To determine the complete mass and radius parameters of the system, we assume a circular orbit. For a binary system, the relationship between the mean density $\bar{\rho}$ of a lobe-filling star and the orbital period follows the equation \citep[chapter 4.4]{2002apa..book.....F}:
\begin{equation}\label{eq14}
\begin{aligned}
\bar{\rho}=\frac{3M_2}{4\pi R_{\rm{L}2}^3}\cong110P_{\mathrm{hr}}^{-2}~\mathrm{g}~\mathrm{cm}^{-3},
\end{aligned}
\end{equation}
where $M_{2}$ represents the mass of the visible star, $R_{\rm{L}2}$ is the effective radius of the Roche lobe, and $P_{\mathrm{hr}}$ denotes the orbital period in the unit of hours. In CVs, the visible star typically fills its Roche lobe, and thus the visible star radius can be approximated by the Roche lobe radius ($R_{\rm{L}2}\approx R_{2}$). The radius of the visible stars $R_{2}$ in J0843 and J0935 is determined through SED fitting, as described in section~\ref{SED}. Using the values of $R_{2}$ and the orbital period $P_{\mathrm{orb}}$, we obtain the masses of the visible stars ($M_{2}$) in J0843 and J0935 to be $0.395 ^{+0.067}_{-0.132}~M_{\odot}$ and $0.108 ^{+0.031}_{-0.030}~M_{\odot}$, respectively.

Given the RV semi-amplitude $K_{\mathrm{abs}}$ of the visible star and the orbital period $P_{\mathrm{orb}}$, the mass function of the unseen companion can be expressed using the following formula:
\begin{equation}\label{eq15}
\begin{aligned}
f(M_1)\equiv\frac{M_1^3\sin^3i}{\left(M_1+M_2\right)^2}=\frac{K_{\mathrm{abs}}^3P_{\mathrm{orb}}}{2\pi G}.
\end{aligned}
\end{equation}
Here, $M_{1}$ is the mass of the unseen companion, $G$ is the gravitational constant, and $i$ is the inclination angle of the orbit. The mass functions for J0843 and J0935 are $0.38 \pm 0.04~M_{\odot}$ and $0.30 \pm 0.01~M_{\odot}$, respectively. Using the mass of the visible star obtained from the period–density relationship (Equation~(\ref{eq14})) and assuming an orbital inclination of 90 degrees, we derive the lower mass limits for the unseen companions of J0843 and J0935 to be $0.83^{+0.05}_{-0.12}~M_{\odot}$ and $0.46 \pm 0.03~M_{\odot}$, respectively. The lower limits of the unseen companion masses for J0843 and J0935 are both significantly greater than the masses of their visible stars. Combining the single-peaked CCF profiles presented in section~\ref{subsec:donor rv} and the fact that $M_1\gg M_2$, the unseen companions in J0843 and J0935 ought to be compact objects. Therefore, the possibility that J0935 is a double M-dwarf system can be ruled out.

   \begin{figure}
   \centering
   \includegraphics[width=12.0cm, angle=0]{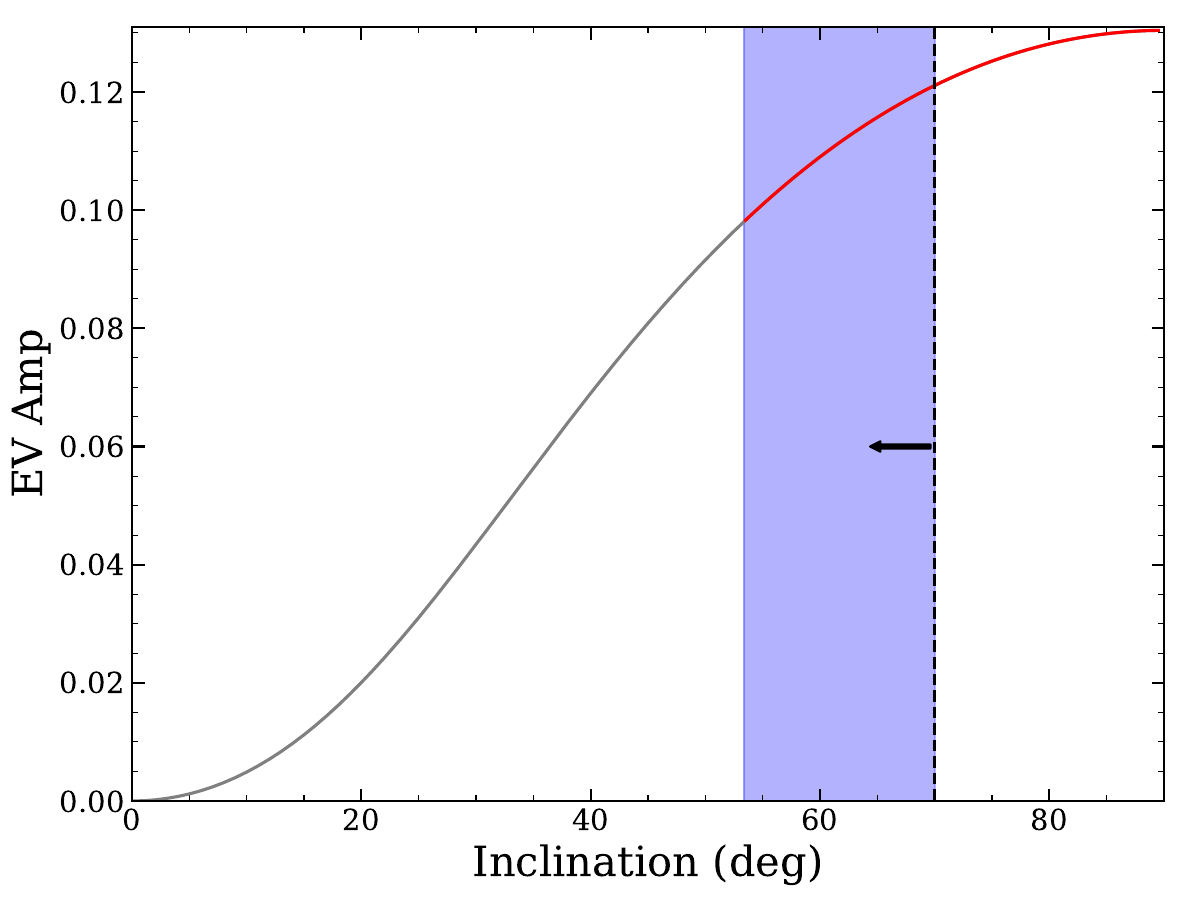}
      \caption{The semi-amplitude of ellipsoidal variability versus orbital inclination. The red curve represents the estimated range of orbital inclinations derived from the amplitudes of ellipsoidal variations, while the black dashed vertical line marks the upper limit of the inclination angle, constrained by the absence of an eclipsing signature in the light curves. The blue region highlights the range of inclination angles for J0843.}
         \label{fig:fig55}
   \end{figure}

In section~\ref{RVEm}, we measured and fitted the RVs of the $\mathrm{H}\mathrm{\alpha}$ emission lines. Our analysis indicates that the $\mathrm{H}\mathrm{\alpha}$ emission lines in J0935 do not solely originate from the accretion disc and thus do not trace the motion of the white dwarf. For J0843, we observe that its emission and absorption lines exhibit opposite velocity trends, leading us to initially hypothesize that the $\mathrm{H}\mathrm{\alpha}$ emission lines may arise from either the accretion disc or the white dwarf, potentially tracing the motion of the white dwarf. It is important to note that the subsequent analysis of J0843 is based on this assumption. Assuming the $\mathrm{H}\mathrm{\alpha}$ emission lines are associated with the motion of white dwarfs, we derive a mass ratio ($q$) of $0.41 \pm 0.03$. Given the visible star mass ($M_{2}$), the mass of the white dwarf is determined to be $0.963^{+0.178}_{-0.329}~M_{\odot}$.

Due to the limited spectral data, the origin of the $\mathrm{H}\mathrm{\alpha}$ emission lines in J0843 remains uncertain. Hence, we independently estimate the white dwarf mass using the mass function to validate the mass derived from the mass ratio. Assuming a system inclination of 90 degrees, we obtain a lower limit of 0.83 $M_{\odot}$ for the white dwarf mass. To determine the true mass, however, an estimate of the system's inclination angle is required. The system inclination of J0843 can be estimated using the amplitudes of ellipsoidal variations. Moreover, the absence of eclipses in the light curve also enables us to establish an upper limit on the inclination.

We adopt the method proposed by \citet{2021MNRAS.501.2822G}, which offers a correction to the analytical approximation of ellipsoidal modulation presented by \citet{1993ApJ...419..344M}. This corrected approximation indicates that the ellipsoidal amplitude depends on the inclination, the mass ratio, and the filling factor, and can be expressed as follows:  
\begin{equation}\label{eq55}
\begin{aligned}
\frac{\Delta L}{\overline{L}}\approx\frac{1}{\overline{L}/L_0}\alpha_2f^3E^3(q')q'\sin^2iC(q',f)\mathrm{cos}2\phi,
\end{aligned}
\end{equation}
where $q'$ denotes the mass ratio defined as $q' \equiv M_1/M_2$, $\phi$ represents the orbital phase, $\overline{L}$ is the average luminosity and $L_0$ is the stellar brightness. Additionally, $\alpha_2$ is a function of the linear limb- and gravity-darkening coefficients of the detected star, $E(q')$ refers to the approximation by \citet{1983ApJ...268..368E}, and $C(q',f)$ is the correction factor related to $q'$ and $f$. In J0843, the visible star obviously fills its Roche lobe, and the filling factor $f$ is constrained to $1$. For a given visible star mass, the mass function imposes additional constraints on the mass ratio and inclination angle. Therefore, by integrating the orbital parameters and stellar properties, the inclination angle and the white dwarf mass can be derived from the amplitude of the ellipsoidal variation. Figure~\ref{fig:fig55} displays the relationship between the semi-amplitude variation and the orbital inclination of J0843.

\begin{table}
\bc
\begin{minipage}[]{100mm}
\caption[]{Orbital and stellar parameters of J0935 and J0843.\label{table:parameters}}\end{minipage}
\setlength{\tabcolsep}{1pt}
\small
 \begin{tabular}{ccccc}
  \hline\noalign{\smallskip}
Parameter & Unit & Value(J0935) & Value(J0843) &Note \\
  \hline\noalign{\smallskip}
R.A.             & h:m:s (J2000) & 09:35:55.08       & 08:43:03.49            & Right Ascension \\
Decl.            & d:m:s (J2000) & +04:29:16.3       & -01:48:58.5            & Declination \\
Gaia parallax    & mas           & 2.76 $\pm$ 0.03   & 0.45 $\pm$ 0.14        & Parallax measured by Gaia EDR3 \\
d(Gaia)          & pc            & $364 ^{+29}_{-26}$  & $2045 ^{+555}_{-414}$       & Distance derived from Gaia EDR3 \\
G-band magnitude & mag           & 18.65 $\pm$ 0.01  & 17.79 $\pm$ 0.01       & G-band magnitude from Gaia EDR3 \\ 
\hline
$P_{\text{orb}}$ & days          & 0.1584(4)        & 0.34615(0)               & Orbital period \\
$T_0$            & HJD           & 2459016.94783      & 2458211.30978          & Ephemeris zero-point \\
$K_{\text{abs}}$ & km s$^{-1}$   & 263 $\pm$ 2     & 219 $\pm$ 7                  & RV semi-amplitude of the visible star \\
$K_{\text{em}}$  & km s$^{-1}$   & 66 $\pm$ 3       & -90 $\pm$ 6                  & RV semi-amplitude of the white dwarf \\
$\gamma_1$       & km s$^{-1}$   & 30 $\pm$ 3        & -27 $\pm$ 2                    & Systemic RV of the visible star \\
$\gamma_2$       & km s$^{-1}$   & 33 $\pm$ 3        & -12 $\pm$ 2                    & Systemic RV of the white dwarf \\ 
\hline
$T_{\text{eff}}$ & K             & $3973^{+174}_{-95}$   & $4643^{+198}_{-166}$      & Effective temperature from SED fitting \\
$\log g$         & dex           & $4.91 \pm 0.06$        & $4.48^{+0.36}_{-0.37}$           & Surface gravity from SED fitting \\
Metallicity      & [M/H]         & $-0.10^{+0.10}_{-0.12}$       & $-0.14^{+0.18}_{-0.14}$     & Metallicity from SED fitting \\
$R_2$          & \(R_{\odot}\)  & $0.271^{+0.024}_{-0.024}$ & $0.704^{+0.038}_{-0.089}$ & Effective radius of the visible star \\
$M_{2}$        & \(M_{\odot}\)  & $0.108 ^{+0.031}_{-0.030}$   & $0.395 ^{+0.067}_{-0.132}$  & Mass of the visible star \\
$q$   & $q \equiv M_2/M_1$  &                       & $0.41 \pm 0.03$& The mass ratio of the visible star to the white dwarf \\
$M_1$          & \(M_{\odot}\)     & $\geq0.46 \pm 0.03$       & $0.963^{+0.178}_{-0.329}$       & Mass of the white dwarf \\
$A(V)$ & mag   & $0.05^{+0.02}_{-0.03}$       & $0.03^{+0.00}_{-0.01}$                           & Extinction from 3D dustmaps \\
  \noalign{\smallskip}\hline
\end{tabular}
\ec
\end{table}

Note that the above analytical model does not consider the luminosity contribution from the accretion disc. If the accretion disc contributes significantly to the total luminosity while exerting a negligible effect on the light variations, the actual ellipsoidal variation amplitude would be larger, leading to a higher inferred inclination angle. The results reveal that the orbital inclination of J0843 lies within the range of $55^\circ - 90^\circ$, as indicated by the red curve in Figure~\ref{fig:fig55}.

On the other hand, the absence of an eclipsing signature in the light curves of J0843 suggests that $R_2 \lesssim a \cos(i)$. For a given inclination angle, the mass of the white dwarf can be derived from Equation~(\ref{eq15}). The binary separation $a$ can be determined by $G(M_1+M_2)P_{\mathrm{orb}}^2=4\pi^2a^3$. Consequently, the orbital inclination is constrained to $i \lesssim 70^\circ$ to satisfy $R_2 \lesssim a \cos(i)$, as shown by the black dashed vertical line in Figure~\ref{fig:fig55}. By combining this with the range of $55^\circ - 90^\circ$, the orbital inclination of J0843 is expected to lie in the range $55^\circ - 70^\circ$. Accordingly, the mass of the white dwarf is estimated to be in the range $0.93 - 1.21~M_{\odot}$, which is consistent with our estimation of $0.963^{+0.178}_{-0.329}~M_{\odot}$ presented in the third paragraph of this section.

\section{Discussion}\label{sec:discuss}
\textit{SDSS J093555.1+042915}~The spectral component of the visible star in J0935 is most prominent, identifying it as a typical M-type star. The Gaia EDR3 distances calculated in \citet{2021yCat.1352....0B} for J0935 is $364 ^{+29}_{-26}$ pc. Using the eFEDS flux value of $6.69 \pm 0.76 \times 10^{-14} \text{erg s}^{-1} \text{cm}^{-2}$, we estimate the X-ray luminosity of J0935 to be $1.06 \times 10^{30} \text{erg s}^{-1}$.
The fit to the absorption line RVs allows for a determination of the orbital period of J0935 as 0.1584 days. The $\mathrm{H}\mathrm{\alpha}$ emission line profile of J0935 varies with phase, particularly exhibiting a pronounced double-peak structure near phases 0.25 and 0.75. Binary systems with two M-dwarf stars can produce such double-peaked variations. However, the single-peaked CCF profile of J0935 and its unseen companion mass lower limit, which exceeds the mass of the visible star, indicate that it is not a binary M-dwarf system. Therefore, we propose that J0935 is still a WD-M dwarf binary, but its $\mathrm{H}\mathrm{\alpha}$ emission likely originates not only from the accretion disc but also from the visible star's stellar activity or its illuminated regions.

\textit{SDSS J084303.5$-$014858}~The spectrum of J0843 exhibits prominent Balmer and He emission lines, characteristic of CV systems. Additionally, the presence of Mg absorption lines in the blue arm of the spectrum implies that the visible star is likely a late K-type star. The Gaia distance of J0843 is $2045 ^{+555}_{-414}$ pc \citep{2021yCat.1352....0B}, and the eFEDS flux is $1.19 \pm 0.14 \times 10^{-13} \text{erg s}^{-1} \text{cm}^{-2}$. The X-ray luminosities (in the 0.2–2.3 keV range) of J0843 is $6.0 \times 10^{31} \text{erg s}^{-1}$. We recognize that the near-infrared data offer the best representation of the companion star's intrinsic properties, as it is less affected by the accretion disc, white dwarf and circumbinary disc. In performing the SED fitting for J0843, we focus on the data points from Pan-STARRS PS1, 2MASS, and WISE $W_{1}$ and $W_{2}$. This choice is critical for constraining the mass of the white dwarf using the radius of the visible star derived from SED fitting. 

\textit{SDSS J091248.24$-$000721}~The periods for J0912 found in ZTF and CRTS are inconsistent, potentially attributed to the limited number of photometric data points available and the fainter magnitudes of the star. The ZTF r-band comprises 121 points, whereas the g-band is only 37 points, and the CRTS provides 66 points. Consequently, we propose 0.0862 days as the potential period for J0912. Due to insufficient SDSS spectra for J0912, a direct search for the period using RVs is also not feasible. We may need to perform follow-up observations to acquire additional spectra to search for its orbital period.

\textit{SDSS J084641.0+021823 }~J0846 exhibits the highest peak at 0.4999 days in both the ZTF r-band and g-band data using the Lomb-Scargle method. Given that ZTF is a ground-based telescope, this period is likely influenced by the Earth's rotation. The period searched in the CRTS data is 0.6566 days. However, it is subject to some uncertainty due to the lack of a prominent peak (Figure~\ref{fig:fig6}). Consequently, the period of J0846 necessitates further investigation.

\section{Summary}\label{sec:summary}
We crossmatched point sources from the eFEDS survey with SDSS-V DR18 data and identified 17 CVs. Utilizing light curves from ZTF and CRTS, we constrained the photometric periods of four sources (J0843, J0846, J0912 and J0935) and obtained the orbital periods of J0843 and J0935 to be 0.34615 days and 0.1584 days by fitting the RVs of the absorption lines. By combining the RV curve amplitudes of the absorption lines and the orbital periods, the mass functions of J0843 and J0935 were determined to be $0.38 \pm 0.04~M_{\odot}$ and $0.30 \pm 0.01~M_{\odot}$, respectively.

For J0935, the RV amplitude of its $\mathrm{H}\mathrm{\alpha}$ emission line and the absorption line had the same direction. This made it impossible to constrain the white dwarf mass using the visible star's mass. Instead, only a lower limit for the white dwarf mass could be derived through the mass function. For J0843, the radius of the visible star was constrained to $R_{2} = 0.704^{+0.038}_{-0.089}~R_{\odot}$ by the results of the SED fitting in the near-infrared band. Using the period-density relationship for Roche lobe-filling stars, we obtained a mass of $M_{2} = 0.395 ^{+0.067}_{-0.132}~M_{\odot}$ for the visible star. Based on the assumption that the $\mathrm{H}\mathrm{\alpha}$ emission lines of J0843 reflect the motion of the white dwarf, the white dwarf mass was determined to be $M_{1} = 0.963^{+0.178}_{-0.329}~M_{\odot}$. This value is consistent with the white dwarf mass range ($0.93 - 1.21~M_{\odot}$) derived from the orbital inclination.

\normalem
\begin{acknowledgements}
We thank Hao-Bin Liu and Qian-Yu An for helpful discussion. We sincerely appreciate the constructive comments and valuable suggestions from the anonymous referee, which have significantly improved this work. This work was supported by the National Key R\&D Program of China under grants 2023YFA1607901 and 2021YFA1600401, the National Natural Science Foundation of China under grants 12433007, 11925301, 12033006, 12221003, and 12263003, and the fellowship of China National Postdoctoral Program for Innovation Talents under grant BX20230020. We acknowledge the science research grants from the China Manned Space Project.

\end{acknowledgements}
  
\bibliographystyle{raa}
\bibliography{ms2025-0230refs}

\end{document}